\documentclass[12pt,a4paper,aps,amsmath]{revtex4}
\usepackage{bm}
\usepackage{graphics}
\usepackage{dcolumn}
\usepackage{enumerate}
\usepackage{latexsym}
\usepackage{feynmf}
\begin{document}

\title{The 2-matrix of the spin-polarized electron gas:
contraction sum rules and spectral resolutions}

\maketitle

\newcommand{\integral}{\mathop{\int d^3r_2}\limits}
\newcommand{\integralp}{\mathop{\int d^3r'}\limits}

P. Ziesche \\
Max-Planck-Institut f\"ur Physik komplexer Systeme, \\
N\"othnitzer Str. 38, D-01187 Dresden, Germany \\
and \\
F. Tasn\'adi \\
Leibniz-Institut f\"ur Festk\"orper- und Werkstoffforschung Dresden, Germany, \\
and University of Debrecen, Hungary \\



{\bf Abstract} \\
The spin-polarized homogeneous electron gas with densities $\rho_\uparrow$ and
$\rho_\downarrow$ for electrons with spin `up' ($\uparrow$) and spin `down'
($\downarrow$), respectively, is systematically analyzed with respect to its
lowest-order reduced densities and density matrices and their mutual relations.
The three 2-body reduced density matrices
$\gamma_{\uparrow\uparrow}$, $\gamma_{\downarrow\downarrow}$, $\gamma_a$ are
4-point functions for electron pairs with spins $\uparrow\uparrow$,
$\downarrow\downarrow$, and antiparallel, respectively. From them, three
functions $G_{\uparrow\uparrow}(x,y)$, $G_{\downarrow\downarrow}(x,y)$,
$G_a(x,y)$, depending on only two variables, are derived. These functions
contain not only the pair
densities according to $g_{\uparrow\uparrow}(r)=G_{\uparrow\uparrow}(0,r)$,
$g_{\downarrow\downarrow}(r)=G_{\downarrow\downarrow}(0,r)$, $g_a(r)=G_a(0,r)$
with $r=|\mathbf{r}_1-\mathbf{r}_2|$, but also the 1-body reduced density
matrices $\gamma_\uparrow$ and $\gamma_\downarrow$ being 2-point functions
according to $\gamma_s=\rho_sf_s$ and $f_s(r)=G_{ss}(r,\infty)$ with $s=\uparrow,
\downarrow$ and $r=|\mathbf{r}_1-\mathbf{r}'_1|$. The contraction properties of
the 2-body reduced
density matrices lead to three sum rules to be obeyed by the three key functions
$G_{ss}$, $G_a$. These contraction sum rules contain corresponding normalization
sum rules as special cases. The
momentum distributions $n_\uparrow(k)$ and $n_\downarrow(k)$, following from
$f_\uparrow(r)$ and $f_\downarrow(r)$ by Fourier transform, are correctly
normalized through $f_s(0)=1$. In addition to the non-negativity conditions
$n_s(k),g_{ss}(r),g_a(r)\geq 0$ [these quantities are probabilities], it holds
$n_s(k)\leq 1$ and
$g_{ss}(0)=0$ due to the Pauli principle and $g_a(0)\leq 1$ due to the Coulomb
repulsion. Recent parametrizations of the pair densities of the
spin-unpolarized homogeneous electron gas in terms of 2-body wave functions
(geminals) and corresponding occupancies are generalized (i) to the
spin-polarized case and (ii) to the 2-body reduced density matrix giving thus
its spectral resolutions. \\

{\bf 1. Introduction} \\
The homogeneous electron gas (HEG) is an important and widely used model for
the phenomenon called electron correlation \cite{Mar,Gon1,Gon}. Its model
parameters are
$\rho_\uparrow$ and $\rho_\downarrow$, the densities of electrons with spin-up
($\uparrow$) and spin-down ($\downarrow$), respectively. Equivalent parameters
are the density $\rho=\rho_\uparrow+\rho_\downarrow$ (from which follows the
density parameter $r_s$ defined by $4\pi r_s^3/3=1/\rho$ in a.u.) and
$\zeta=(\rho_\uparrow-\rho_\downarrow)/\rho$, the spin-polarization. $r_s$ is
the (Wigner) radius of a sphere containing on average one electron. It measures
simultaneously the coupling or interaction strength. The spin-dependent Wigner
radii $r_{s\uparrow}$ and $r_{s\downarrow}$ are analogously defined by
$4\pi r_{s\uparrow}^3/3=1/\rho_\uparrow$ and $4\pi r_{s\downarrow}^3/3=
1/\rho_\downarrow$, respectively. This (jellium)
model is relevant for the understanding of many effects in simple metals and
semiconductors and it plays a crucial role in
providing input quantities for approximate approaches to the many-electron
problem of nonuniform density in solid-state theory and quantum chemistry. So
the local-spin density approximation (LSD) of the density functional theory
relies on $e(r_s,\zeta)$, the energy per particle of the HEG. Corresponding LSD
(and beyond-LSD) functionals $E_{\rm xc}[\rho_\uparrow,\rho_\downarrow]$ often
treat the electron-electron interaction as a whole, without
splitting it into its spin-resolved ($\uparrow\uparrow, \downarrow\downarrow$,
and `$a$' for spin-antiparallel) components. But some of the most popular
functionals are constructed by considering spin-resolved contributions to the
correlation energy. This makes the recent interest in the spin-polarized HEG
understandable. The system under consideration has the two afore mentioned
model parameters $r_s$ and $\zeta$. Its ground state is characterized by the
reduced densities $n_\uparrow(k)$, $n_\downarrow(k)$ [the momentum
distributions] and $g_{\uparrow\uparrow}(r)$, $g_{\downarrow\downarrow}(r)$,
$g_a(r)$ [the pair densities (PDs)]. Corresponding energies are $t_\uparrow$,
$t_\downarrow$ [the kinetic energies] and $v_{\uparrow\uparrow}$,
$v_{\downarrow\downarrow} $, $v_a$ [the interaction
energies] such that the total bulk energy $e$ is given by
$e=e_\uparrow+e_\downarrow$ where $e_\uparrow= t_\uparrow+v_\uparrow$ and
$e_\downarrow=t_\downarrow+v_\downarrow$ with $v_\uparrow=v_{\uparrow\uparrow}+
v_a$ and $v_\downarrow=v_{\downarrow\downarrow}+ v_a$. All these quantities
depend parametrically on $r_s$ and $\zeta$. The PDs have been studied recently
in Refs. \cite{Cep,Per,Ort,Gor1,Gor2}. Recent studies of $n(k)$ for the
spin-unpolarized case are in Ref. \cite{Zie}. The total (spin-weighted) PD
$g=(\rho_\uparrow/\rho)^2g_{\uparrow\uparrow}+
(\rho_\downarrow/\rho)^2g_{\downarrow\downarrow}+
2(\rho_\uparrow\rho_\downarrow/\rho^2)g_a$ is related to the total
momentum distribution $n(k)=\frac{1}{2}[n_\uparrow(k)+n_\downarrow(k)]$ (i) by
the virial
theorem \cite{Mar2,Mac} and (ii) through the 2-body reduced density matrix
(2-matrix) \cite{Davi,Cios,Col1,Davi2}. The latter means: its diagonal elements
give the PDs and its contraction yields the 1-matrix, which is the Fourier
transform of the momentum distribution. \\

These mutual relations between the PDs and the momentum distributions are
investigated for the spin-polarized
HEG in the following. In detail its 2-matrix and
quantities derived from it are systematically analyzed in particular in terms of
the already mentioned contraction sum rules (SRs) and the so-called spectral
resolutions. This more general analysis (with introductory parts in Secs. 2-6
and the main parts in Secs. 7-9)
includes the special cases of `no interaction' (but $\zeta\neq 0$), `no
polarization' (but $r_s\neq 0$), and `full polarization' (with $\zeta=1$ and
$r_s\neq 0$) [they are summarized in Apps. 1, 2, 3, respectively]. Thereby four
questions are considered: \\
1. For the spin-unpolarized HEG it is well-known as already mentioned, that the
combined virial and
Hellmann-Feynman theorem relates the interaction energy $v$ to the kinetic
energy $t$, such that $t(r_s)$ can be calculated provided $v(r_s)$ is known from
the PD $g(r;r_s)$. The question arises whether there are theorems for the
spin-polarized HEG, such that not only $t$, but also its spin-components
$t_\uparrow$ and $t_\downarrow$ can be calculated from $v_{\uparrow\uparrow}$,
$v_{\downarrow\downarrow}$, $v_a$, which are available because
$g_{\uparrow\uparrow}$, $g_{\downarrow\downarrow}$, $g_a$ are known as
functions of $r$ and $r_s,\zeta$ \cite{Cep,Per,Ort,Gor1,Gor2}. The answer in
Sec. 3 is `no' (disagreeing with a formula in Ref. \cite{Cac}).  \\
2. The virial theorem is an integral relation between $n(k)$ and $g(r)$. Other
relations arise from their common origin, namely the 2-body reduced density
matrix. The question `how these relations simplify themselves for homogeneous
systems ?' is answered in Sec. 8. \\
3. Kimball \cite{Kim1} and Overhauser \cite{Over} introduced a parametrization
of the HEG-PD in terms of 2-body wave functions (so-called geminals), which are
the solutions of a 2-body Schr\"odinger equation with an appropriately chosen
effective interaction (screened Coulomb repulsion), which describes the
scattering of two electrons in the medium of all the others. Further
developments of this idea \cite{Gor3,Dav} allows one to calculate
the PD in excellent
agreement with the results of quantum Monte Carlo calculations \cite{Ort,Cep}.
Because nowadays the PDs not only for $\zeta=0$ but also for $\zeta\neq 0$ are
available \cite{Ort,Gor1,Gor2}, the question may be posed
whether in this more general spin-polarized case the PDs can be parametrized
again in terms of geminals following from a screened Coulomb repulsion. In
Sec. 9 the answer `yes' is assumed and a corresponding analysis is presented as
the necessary pre-stage of a numerical study. \\
4. In Ref. \cite{ZiePR} it has been shown how the PD-normalization can be
expressed in terms of scattering phase shifts of the geminals similar as this
is known from the Friedel sum rule. Moreover in Ref. \cite{Ziepss} the
Kimball-Overhauser approach is generalized with the assumption that the
PD-geminals can be used to represent not only the PD but also the 2-matrix. The
contraction of which gives the 1-matrix and thus the momentum distribution.
Along this line the normalization SR of Ref. \cite{ZiePR} is generalized in Ref.
\cite{Ziepss} to a contraction SR, which allows one in
principle to self-consistently calculate the momentum distribution provided the afore mentioned geminals are known. For a summary cf. Ref. \cite{Ann}. Again
the question arises whether this analysis can be extended to non-vanishing
spin-polarization. Preliminary considerations are presented in Sec. 10. \\

The paper is structured as it follows. Sec. 1 defines the HEG by its
Hamiltonian and lists the ground state quantities. In Sec. 2 the virial and
Hellmann-Feynmam theorems are revisited. Secs. 4 and 5 present the 1-matrix
$\gamma_1$ and the 2-matrix $\gamma_2$, respectively, with the spin-structure
of $\gamma_2$, derived in App. 4. Sec. 6 summarizes the four PDs and their
normalizations. Sec. 7 describes how the 1-matrix can be obtained from the
2-matrix by contraction. Apps. 1, 2, 3 summarize the special cases of `no
interaction', `no spin-polarization', `full spin-polarization', respectively.
In Sec. 8 the 2-matrix is discussed in terms of Wick's
theorem and size-extensivity, guaranteed by linked diagrams and cumulants. Sec.
9 presents the spectral resolutions of the 2-matrix and the cumulant 2-matrix
in terms of geminals and corresponding occupancies. Sec. 10 expresses the
normalization and contraction SRs in terms of the scattering phase shifts which
describe the asymptotics of the geminals.  \\

{\bf 2. The system and its ground state properties} \\
The GS of the spin-polarized HEG is characterized by the average densities
$\rho_\uparrow$ and $\rho_\downarrow$ of the spin-up, respectively spin-down
electrons or equivalently by the average spin-summed density $\rho=
\rho_\uparrow+\rho_\downarrow$ and the spin-polarization $\zeta=(\rho_\uparrow-
\rho_\downarrow)/\rho$, thus $\rho_\uparrow/\rho=(1+\zeta)/2$ and
$\rho_\downarrow/\rho=(1-\zeta)/2$. Essential parts of the phenomenon `electron
correlation' are contained in the two dimensionless  momentum distributions
$n_\uparrow(k)$, $n_\downarrow(k)$ and in four dimensionless PDs, namely
$g_{\uparrow\uparrow}(r)$, $g_{\downarrow\downarrow}(r)$, and
$g_{\pm}(r)$ or equivalently $g_a=(g_{+}+g_{-})/2$, $\Delta g=(g_{+}-
g_{-})/2$. The index `$\pm$' stands for singlet/triplet, `$a$' for
spin-antiparallel, respectively. To get these GS quantities one has to solve
\begin{equation}
{\hat H}\Psi=E\Psi, \quad {\hat N}_s\Psi=N_s\Psi
\label{schroe}
\end{equation}
with $\hat H=\hat T+\hat V$ and
\begin{eqnarray}
\hat T=\int d1 \; \hat \psi_1^\dag t_1\hat \psi_1 &,& \quad
t_1=-\frac{\hbar^2}{2m}\left (\frac{\partial}{\partial \mathbf{r}_1}\right )^2,\quad\qquad \nonumber \\
\hat V=\int \frac{d1d2}{2!}\;
[:({\hat \rho}_1-\rho_{\Omega 1})v_{12}({\hat \rho}_2-\rho_{\Omega 2}):]\; &,&
\quad
v_{12}=\frac{\epsilon^2}{r_{12}}, \quad r_{12}=|\mathbf{r}_1-\mathbf{r}_2|,
\nonumber \\
\hat \rho_1=\hat \psi_1^\dag\hat \psi^{}_1,\quad
:\hat \rho_1\hat \rho_2:=
\hat \psi_1^\dag\hat \psi_2^\dag\hat \psi^{}_2\hat \psi^{}_1
&,& \quad
\rho_{\Omega 1}=\rho_{s_1}\Theta_{\Omega}(\mathbf{r}_1)
\label{ham}
\end{eqnarray}
and to take the TDL $N_s, \Omega \to \infty$, $N_s/\Omega={\rm const}=
\rho_s$ with $N=N_\uparrow+N_\downarrow$. Here $1=(\mathbf{r}^{}_1,\sigma_1)$,
$\sigma$ and $s=\; \uparrow$ or $\downarrow$, and $\hat{\psi}_1=
\hat{\psi}_{s_1}(\mathbf{r}^{}_1)$.
$\hat{\psi}_s^\dag(\mathbf{r})$ and $\hat{\psi}_s(\mathbf{r})$ are the Fermion
creation and annihilation
field operators such that $\hat \rho_s(\mathbf{r})=\hat \psi_s^\dag(\mathbf{r})
\hat \psi_s(\mathbf{r})$ is
the density operator for electrons with spin $s$ and $\hat \rho(\mathbf{r})=
\hat \rho_\uparrow(\mathbf{r})+\hat \rho_\downarrow(\mathbf{r})$ is the
spin-summed density operator. Hence, $\hat{N}_s=
\int d^3r\  \hat{\psi}^\dag_s(\mathbf{r}) \hat{\psi}^{}_s(\mathbf{r})$
and $\hat{N}=\hat{N}_{\uparrow}+\hat{N}_{\downarrow}$ are the particle-number
operators for spin $s$ and spin-summed, respectively. The operators
$(:\hat\rho_1\hat \rho_2:)$, $\hat \rho_1(\rho_\Omega)_2+
(\rho_\Omega)_1\hat \rho_2$, and
$(\rho_\Omega)_1(\rho_\Omega)_2$ describe the electron-electron ($--$),
electron-background ($-+$ and $+-$), and background-background ($++$)
interaction, respectively.
Corresponding expectation values are the spin-density $\rho_s(\mathbf{r})=
\langle\hat \rho_s(\mathbf{r})\rangle$
for $s=\uparrow,\downarrow$, as well as the two 1-matrices
$\rho_sf_s(|\mathbf{r}_1-\mathbf{r}'^{}_1|)=\langle\psi_s^\dag(\mathbf{r}'_1)
\psi_s(\mathbf{r}_1)\rangle$ and the four dimensionless PDs $g_{ss},g_a,
\Delta g$ defined by
$\rho_s^2g_{ss}(r_{12})=
\langle:\hat \rho_s(\mathbf{r}_1)\hat \rho_s(\mathbf{r}_2):\rangle , \quad
\rho_\uparrow\rho_\downarrow g_a(r_{12})=
\langle\hat \rho_\uparrow(\mathbf{r}_1)\hat
\rho_\downarrow(\mathbf{r}_2)\rangle$, and $\rho_\uparrow\rho_\downarrow
\Delta g(r_{12})=\langle\hat \psi_\downarrow^\dag(\mathbf{r}_1)
\hat \psi_\uparrow^\dag(\mathbf{r}_2)\hat\psi_\downarrow(\mathbf{r}_2
\hat\psi_\downarrow(\mathbf{r}_1)\rangle$. Then the spin-weighted PD
$g(r)$ is defined by $\rho^2g(r)=\rho_\uparrow^2g_{\uparrow\uparrow}(r)+
\rho_\downarrow^2g_{\downarrow\downarrow}(r)+
2\rho_\uparrow\rho_\downarrow g_a(r)$. The momentum distributions are given by
$n_s(k)=\langle\hat a_{s\mathbf{k}}^\dag\hat a_{s\mathbf{k}}\rangle$ and
$n(k)=\frac{1}{2}[n_\uparrow(k)+n_\downarrow(k)]$ with
$\sum_\mathbf{k}n_s(k)=N_s$ and $2\sum_\mathbf{k}n(k)=N$; the normalizations
per particle are $\frac{1}{N}\sum_\mathbf{k}n_\uparrow(k)=\frac{1}{2}(1+\zeta)>
\frac{1}{2}$, $\frac{1}{N}\sum_\mathbf{k}n_\downarrow(k)=\frac{1}{2}(1-\zeta)
<\frac{1}{2}$, and $\frac{2}{N}\sum_\mathbf{k}n(k)=1$. \\

{\bf 3. The virial and Hellmann-Feynman theorems} \\
\noindent
From the five mentioned reduced densities $n_\uparrow(k),n_\downarrow(k),
g_{\uparrow\uparrow}(r), g_{\downarrow\downarrow}(r), g_a(r)$ follow five
contributions to the total energy $E=T+V$
with $T=T_\uparrow+T_\downarrow$ and $V=V_{\uparrow\uparrow}+
V_{\downarrow\downarrow}+2V_a$, being the GS expectation values of $\hat H=
\hat T+\hat V$ with $\hat T={\hat T}_\uparrow+{\hat T}_\downarrow$ and
${\hat V}={\hat V}_{\uparrow\uparrow}+{\hat V}_{\downarrow\downarrow}+
2{\hat V}_{a}$. Now, the virial theorem is one relation between $T$ and $V$ with
the consequence: if $V$ is known, then $T$ can be calculated. The question
arises: Are there perhaps two virial theorems, one for spin $\uparrow$ and one
for spin $\downarrow$, which - for given PDs $g_{ss}$  and $g_a$ - would
possibly allow one to calculate the spin components $T_\uparrow$ and
$T_\downarrow$, separately ? The answer is `no'. To show this, one may define
${\hat H}_s={\hat T}_s+{\hat V}_s$, ${\hat V}_s={\hat V}_{ss}+{\hat V}_a$ and
$E_s=T_s+V_{s}$, $V_s=V_{ss}+V_a$, for $s=\uparrow, \downarrow$, such that
$\hat H={\hat H}_\uparrow+{\hat H}_\downarrow$ and
$E=E_\uparrow+E_\downarrow$, but there is only one virial theorem, to be derived
from the hypervirial theorem $\langle[\hat F,\hat H]\rangle=0$, which holds for
an eigenstate $\Psi$ of $\hat H$ and an arbitrary operator $\hat F$ of the
system; the commutator is denoted by $[A,B]=AB-BA$. With the virial operator
$\hat F=\hat {\mathbf{r}\mathbf{p}}$ and with its spin resolution
${\hat {\mathbf{rp}}}=({\hat {\mathbf{rp}}})_\uparrow+({\hat {\mathbf{rp}}
})_\downarrow$,
$({\hat {\mathbf{rp}}})_s=\int d^3r \, \hat{\psi}_s^\dag(\mathbf{r})\hat{\mathbf{rp
}}
\hat{\psi}_s(\mathbf{r})$ it is easy to show
\begin{eqnarray}
0  =  -\frac{\rm i}{\hbar}\langle[\hat {\mathbf{rp}},\hat H ]\rangle
  & = & \langle [\hat {\mathbf{p}\frac{\partial}{\partial \mathbf{p}}},
\hat T ]\rangle+
\langle[\hat {(-\mathbf{r}\frac{\partial}{\partial \mathbf{r}})},
\hat V ]\rangle \nonumber \\
  & = & 2T+V+3\Omega\langle\frac{\partial \hat V}{\partial \Omega}\rangle, \quad
\frac{\partial \hat V}{\partial \Omega}=\frac{\partial \hat H}{\partial \Omega}.
\label{virial1}
\end{eqnarray}
Only ${\hat V}^{+-}+{\hat V}^{-+}+V^{++}$ depend on the system parameter
$\Omega=$ volume. And there is only one Hellmann-Feynman theorem \cite{Hell}
\begin{equation}
\frac{\partial E}{\partial \Omega}=\langle\frac{\partial \hat H}{\partial \Omega
}\rangle .
\label{Hell}
\end{equation}
The reason: There is only one MB Schr\"odinger equation $(\hat H-E)\Psi=0$.
In addition to the virial theorem (\ref{virial1}) one may derive the relations,
cf. \cite{Isi}
\begin{equation}
0=\langle[(\hat {\mathbf{r}\mathbf{p}})_\uparrow,\hat H]\rangle= 2T_\uparrow+
V_{\uparrow\uparrow}+X+3\Omega\langle\frac{\partial}{\partial \Omega}
\hat H_\uparrow\rangle ,
\label{virialup}
\end{equation}
\begin{equation}
0=\langle[(\hat {\mathbf{r}\mathbf{p}})_\downarrow,\hat H]\rangle=
2T_\downarrow+V_{\downarrow\downarrow}-X+
3\Omega\langle\frac{\partial}{\partial \Omega}\hat H_\downarrow\rangle ,
\label{virialdown}
\end{equation}
where the expection value of
\begin{eqnarray}
\hat X=\frac{1}{2}\int d^3r_1d^3r_2 \, \hat \rho_\uparrow(\mathbf{r}_1)
\hat \rho_\downarrow(\mathbf{r}_2)
\frac{r_1^2-r_2^2}{r_{12}^2}\frac{\epsilon^2}{r_{12}}.
\label{deltaX}
\end{eqnarray}
appears. In Ref. \cite{Isi} the modified (`filtered' with the prefactor
$\mathbf{R}\mathbf{r}/r^2=R\xi/r$, where $\mathbf{R}$ is the centre-of-mass
and $\mathbf{r}=\mathbf{r}_1-\mathbf{r}_2$) Coulomb
repulsion in Eq.(\ref{deltaX}) is referred to as `spin polarization potential'.
Note $\langle\partial \hat H_s/\partial \Omega\rangle\neq\partial E_s/
\partial \Omega$: The HF theorem (\ref{Hell}) does not split into two different
theorems for spin $\uparrow$ and spin $\downarrow$. So, the sum of
Eqs.(\ref{virialup}) and (\ref{virialdown}) yields correctly the VT
(\ref{virial1}), but their difference gives
\begin{equation}
2\Delta T+\Delta V +2X+
3\Omega\langle\frac{\partial}{\partial \Omega}\Delta \hat H\rangle=0
\label{spinvirial}
\end{equation}
with $\Delta T=T_\uparrow-T_\downarrow$, $\Delta V=V_{\uparrow\uparrow}-
V_{\downarrow\downarrow}$ ($V_a$ cancels), $\Delta \hat H=
\hat H_\uparrow-\hat H_\downarrow=\Delta \hat T+\Delta \hat V$. \\

Specification of $\Omega$: For an atom with a pointlike positive background
charge, i.e. $\Omega\to 0$, $\rho\to\infty$, $\rho\Omega\to Z$ (atomic limit),
all the terms with $\partial/\partial
\Omega$ are no longer present. Then the virial theorem (\ref{virial1}) is simply
$2T+V=0$ and the other relation (\ref{spinvirial}) is $2\Delta T+\Delta V+
2X=0$, in Ref. \cite{Isi} called `spin virial theorem'. Contrary to this for
the spin-polarized HEG in its TDL with $N,\Omega\to\infty$ and $t=T/N,v=V/N,
e=E/N$, and $3\Omega\partial e/\partial \Omega=
r_s\partial e/\partial r_s$, from the theorems (\ref{virial1}) and
(\ref{Hell}) follows the bulk virial theorem \cite{Mar2}
\begin{equation}
2t+v+r_s\frac{\partial}{\partial r_s}e=0 \quad {\rm or} \quad
\frac{1}{r_s}\frac{\partial}{\partial r_s}r_s^2t+
\frac{\partial}{\partial r_s}r_sv=0 ,
\label{virial2}
\end{equation}
where the derivative is related to the pressure of the
system: $r_s\partial e/\partial r_s=-3p/\rho$. For the case `no interaction' the
virial theorem (\ref{virial2}) simply says $t^{0}\sim 1/r_s^2$.
Also the surface virial theorem $2t_{\rm surf}+v_{\rm surf}+
(r_s\partial/\partial r_s+2)e_{\rm surf}=0$ can be derived \cite{Vann}. But
what results from Eq.(\ref{spinvirial})? One may write the sum and the
difference of the last terms on the rhs's of Eqs.(\ref{virialup}) and
(\ref{virialdown}) as
\begin{equation}
3\Omega\langle\frac{\partial}{\partial \Omega}\hat H\rangle=
3\Omega\frac{\partial}{\partial \Omega} E -
3\Omega\left (\frac{\partial}{\partial \Omega}\right )_\Psi\langle \hat H\rangle
\label{sum}
\end{equation}
and
\begin{equation}
3\Omega\langle\frac{\partial}{\partial \Omega}\Delta \hat H\rangle=
3\Omega\frac{\partial}{\partial \Omega} \Delta E -
3\Omega\left (\frac{\partial}{\partial \Omega}\right )_\Psi
\langle \Delta \hat H\rangle ,
\label{diff}
\end{equation}
respectively. $(\partial/\partial\Omega)_\Psi$ means differentiation with
respect to $\Psi$, not $\hat H$. Now, the last term on the rhs of
Eq.(\ref{sum}) vanishes,
because of the Hellmann-Feynman theorem (\ref{Hell}), thus giving the bulk
virial theorem (\ref{virial2}). Contrary to this the last term on the rhs of
Eq.(\ref{diff}) gives an expression,
which does not vanish. So, Eq.(\ref{diff}) used in
Eq.(\ref{spinvirial}) gives only the trivial identity $0=0$. This is seen from
\begin{equation}
2\Delta t+\Delta v+r_s\frac{\partial \Delta e}{\partial r_s}=
\frac{1}{r_s}\frac{\partial}{\partial r_s}r_s^2\Delta t+
\frac{\partial}{\partial r_s}r_s\Delta v=
r_s\left (\frac{\partial}{\partial r_s}\right )_\Psi
\langle \frac{\Delta \hat H}{N}\rangle
\end{equation}
and (in the TDL)
\begin{equation}
x=\frac{X}{N}=\frac{\rho_\uparrow\rho_\downarrow}{\rho^2}
2\int d^3r\; \rho \int\frac{d^3R}{\Omega} \frac{R}{r}
\int_{-1}^{+1}\frac{d\xi}{2}\xi
g_\Omega(R,r,\xi)
\frac{\epsilon^2}{r}=O(N^{-2/3}),
\end{equation}
where $\rho_\uparrow\rho_\downarrow g_\Omega(R,r,\xi)=\langle
\hat \rho_\uparrow(\mathbf{r}_1)\hat \rho_\downarrow(\mathbf{r}_2)\rangle$,
$\mathbf{R}=(\mathbf{r}_1+\mathbf{r}_2)/2$,
$\mathbf{r}=(\mathbf{r}_1-\mathbf{r}_2)$,
$\xi=<\hspace{-1.3mm}) (\mathbf{R},\mathbf{r})$, and
$g_\infty(R,r,\xi)=g_a(r)$.
So, from the HEG-PDs $g_{\uparrow\uparrow}$, $g_{\downarrow\downarrow}$, $g_{a}$
as functions of $r$ and the parameters $r_s,\zeta$ one can calculate only
$t$ as a function of $r_s$ and $\zeta$, but not its spin components
$t_\uparrow,t_\downarrow$, separately. Thus Eq.(3.11) of Ref. \cite{Cac}
is wrong. But one may define a quantity $y$ as a function of $r_s$ and
$\zeta$ by
\begin{equation}
2\Delta t+\Delta v+r_s\frac{\partial \Delta e}{\partial r_s}=y
\end{equation}
(so this is not a theorem !). For `no interaction' it is $\Delta v^{0}=0,
\Delta t^{0}\sim 1/r_s^2$, so $y$ vanishes with $r_s\to 0$.  - In the
literature there are hints on a magnetic virial theorem if one adds to the
Hamiltonian (\ref{ham}) a Zeeman term $\sim \hat S_z=\frac{\hbar}{2}
(\hat N_\uparrow-\hat N_\downarrow)$ \cite{Cac,Qian}. \\

{\bf 4. The 1-matrices} \\
The 1-matrix is defined by $\gamma_1(1|1')=\langle \hat{\psi}^+_{1'}
\hat{\psi}^{}_1\rangle$, normalized as
$\text{Tr}\gamma_1= N_{\uparrow}+N_{\downarrow}=N$.
In terms of natural spin-orbitals $\psi_k(1)$ and corresponding occupation
numbers $\nu_k$ the spectral resolution of $\gamma_1$ is $\gamma_1(1|1')=
\sum_k\psi_k(1)\nu_k\psi_k^*(1')$ with $0<\nu_k<1$ and $\sum_k\nu_k=N$.
For a spin-independent Hamiltonian as $\hat H$ of Eqs.(\ref{schroe}) and
(\ref{ham}) it is $\psi_k(1)=\delta_{s,\sigma_1}\varphi_{\mathbf{k}}
(\mathbf{r}_1)$ with $\langle \psi_\uparrow^\dag
(\mathbf{r}')\psi_\downarrow(\mathbf{r})\rangle=0$ and $\langle
\psi_\downarrow^\dag(\mathbf{r}')\psi_\uparrow(\mathbf{r})\rangle=0$. Thus
\begin{equation}
\gamma_1(1|1')= \delta_{\sigma^{}_1,\sigma'_1}[\delta_{\sigma_1,\uparrow}
\gamma_{\uparrow}(\mathbf{r}^{}_1|\mathbf{r}'_1)+\delta_{\sigma_1,\downarrow}
\gamma_{\downarrow}(\mathbf{r}^{}_1|\mathbf{r}'_1)]
\label{onematrix}
\end{equation}
with
\begin{equation}
\gamma_s(\mathbf{r}^{}|\mathbf{r}')=
\sum_{\mathbf{k}}\varphi_{\mathbf{k}}(\mathbf{r})n_s(\mathbf{k})
\varphi_{\mathbf{k}}^*(\mathbf{r}'),
\text{Tr}\gamma_s=\sum_{\mathbf{k}}n_s(\mathbf{k})=N_s,
\sum_{\mathbf{k}}=\int\frac{\Omega d^3k}{(2\pi)^3}.
\end{equation}
The spin-traced 1-matrix
is $\gamma=\gamma_\uparrow+\gamma_\downarrow =\rho_\uparrow f_\uparrow(r)+
\rho_\downarrow f_\downarrow(r)$. \\

For a homogeneous system the natural orbitals are plane waves
$\varphi_{\mathbf{k}}(\mathbf{r})=
\frac{1}{\sqrt \Omega}{\rm e}^{{\rm i}\mathbf{k}\mathbf{r}}$ and it is
$\gamma_s(\mathbf{r}|\mathbf{r}')=
\rho_s f_s(|\mathbf{r}-\mathbf{r}'|)$, which defines the dimensionless
1-matrices
\begin{equation}
f_s(r)=\frac{1}{N_s}
\sum_{\mathbf{k}}n_s(k){\rm e}^{{\rm i}\mathbf{k}\mathbf{r}},
\label{1matrix}
\end{equation}
displayed in Fig. 1. The Fourier transforms of $f_s(r)$ give the momentum
distributions $n_s (k)$ with
$0\leq n_s(k)\leq 1$ and with a large and a small Fermi sphere:
$(k_{{\rm F}s})^3=6\pi^2\rho_s$, $k_{{\rm F}\uparrow}> k_{{\rm F}\downarrow}$.
With $k_{\rm F}^3=3\pi^2\rho$ it is $k_{\rm F \uparrow}/k_{\rm F}=
(1+\zeta)^{1/3}$ and $k_{\rm F \downarrow}/k_{\rm F}=(1-\zeta)^{1/3}$.  $f_s(0)
=1$ makes the $n_s(k)$ correctly normalized: $\sum_{\mathbf{k}}n_s(k)=N_s$.
The deviations of $n_s(k)$ from 0 and 1 (its non-idempotency) measure the
correlation strength: $\sum_\mathbf{k}[n_s(k)]^2<N_s$. The spin-summed momentum
distribution is $n(k)=\frac{1}{2}[n_\uparrow(k)+n_\downarrow(k)]$. Whether the
asymptotics
of the $n_s(k)$ is spin-independent has to be checked. If so, it should hold
$n_s(k\to \infty)=(2\alpha_0/3\pi)^2g_a(0)/k^8$, $\alpha_0=(4/9\pi)^{1/3}r_s$
\cite{Kim2,Yas}. \\

For the special cases `no interaction', `no polarization', and `full
polarization' cf. Apps. 1, 2, and 3, respectively. \\

The recently available momentum distribution $n(k,r_s)$ of the spin-unpolarized
HEG ($k$ measured in units of $k_{\rm F}$) \cite{Zie} can be used to
approximately
construct the momentum distributions of the spin-polarized HEG for small values
of $\zeta$:
\begin{equation}
n_\uparrow (k,r_s,\zeta)\approx
n\left (\frac{k}{(1+\zeta)^{1/3}},
\frac{r_s}{(1+\zeta)^{1/3}}\right ), \quad
n_\downarrow(k,r_s,\zeta)\approx
n\left (\frac{k}{(1-\zeta)^{1/3}},
\frac{r_s}{(1-\zeta)^{1/3}}\right ).
\label{approx}
\end{equation}
This allows one to calculate the above mentioned spin components
$t_\uparrow(r_s,\zeta)$ and $t_\downarrow(r_s,\zeta)$ of the total kinetic
energy $t(r_s,\zeta)=t_\uparrow(r_s,\zeta)+t_\downarrow(r_s,\zeta)$ and to
control it by comparing with $t(r_s,\zeta)$, calculated from the PDs with the
help of the virial theorem (\ref{virial2}). Results for $\zeta=1/3$ and $r_s<10$
are displayed in Figs. 2 and 3. They show that the approximation (\ref{approx})
affords the correct $t(r_s,\zeta)$.

Related quantities derived from $f_\uparrow$, $f_\downarrow$ or $n_\uparrow$,
$n_\downarrow$ are the 1-matrices (referred to as non-idempotency matrices)
\begin{eqnarray}
\beta_{s}(r_{11'})=
\left (\frac{\rho_s}{\rho}\right )^2\int d^3r_2\rho f_s(r_{12})f_s(r_{21'})=
\frac{1}{N}\sum_{\mathbf{k}}[n_s(k)]^2{\rm e}^{i \mathbf{k}\mathbf{r}_{11'}}&,&
\quad \beta_{s}(0)=b_{s}, \nonumber \\
\beta_a(r_{11'})=\frac{\rho_\uparrow\rho_\downarrow}{\rho^2}\int d^3r_2
\rho f_\uparrow(r_{12})f_\downarrow(r_{21'})=
\frac{1}{N}\sum_{\mathbf{k}}n_\uparrow(k)
n_\downarrow(k){\rm e}^{i \mathbf{k}\mathbf{r}_{11'}}&,& \quad \beta_a(0)=b_a.
\label{beta}
\end{eqnarray}
The product $n_\uparrow(k)n_\downarrow(k)$ makes the $\mathbf{k}$-summation to
run over three different regions: The inner sphere $k<(1-\zeta)^{1/3}$, the
shell $(1-\zeta)^{1/3}<k<(1+\zeta)^{1/3}$, and the outer region
$k>(1+\zeta)^{1/3}$. The quantities $b_{s}$ and $b_a$ are related to
spin-generalized L\"owdin parameters
\begin{eqnarray}
c_{s}=1-\frac{N}{N_s}b_{s}&=&1-\frac{1}{N_s}\sum_{\mathbf{k}}[n_s(k)]^2,
\nonumber \\
c_a=1-\frac{N}{N_\downarrow}b_a&=&1-
\frac{1}{N_\downarrow}\sum_{\mathbf{k}}n_\uparrow(k)n_{\downarrow}(k),
\label{Low}
\end{eqnarray}
which vanish for `no interaction', cf. App. 1. For `no polarization' (cf. App.
2) it becomes $n_\uparrow(k)=n_\downarrow(k)\equiv n(k)$ and therefore $b_{s}=
b_a\equiv b$, $c_{s}=c_a\equiv c$, $c=1-2b$, and $b=\frac{1}{N}
\sum_{\mathbf{k}} [n(k)]^2$. It was already P.-O. L\"owdin who has asked for the
meaning of ${\rm Tr}\gamma_1^2$. The answer is nowadays: $1-{\rm Tr}\gamma_1^2
/N$ is the normalization of the cumulant 2-matrix and one possible index of the
correlation strength. As seen from $c=\frac{2}{N}\sum_\mathbf{k}
n(k)[1-n(k)]$, this expression is particle-hole symmetric \cite{Rus}. In
addition to the above defined non-idempotency matrices, the 1-matrices
$\alpha(r)$, $\tilde \alpha(r)$ appear,
when contracting the 2-matrix, cf. Eq.(\ref{contractionanti}). They cause (for spin-antiparallel electron pairs) the difference between singlet and triplet PDs
and describe fluctuations, which mimick spin-flip processes. This means the
following: One may consider a spatial part (region, domain, or fragment)
$\omega$ of the
system, say a sphere, which contains on average $\omega \rho$ electrons. One may
furthermore ask for the probability of finding $N_\omega\neq\omega\rho$
electrons in $\omega$. This can be realized by a certain number
$N_{\omega\uparrow}$ of spin-up electrons and a certain number
$N_{\omega\downarrow}$ of spin-down electrons. These numbers can change under
the constraint $N_{\omega\uparrow}+N_{\omega\downarrow}=N_\omega$, what looks
effectively as if electrons in $\omega$ had made spin-flips, although such
fluctuations are based only on electrons (with spin-up and spin-down) leaving
or entering the region $\omega$. In general, for particle number fluctuations in
spatial regions cf. Ref. \cite{Zie1}. - For `no interaction' it is
$\alpha^{0}(r)=\frac{N_\downarrow}{N}f_\downarrow^{0}(r)$
and $\tilde \alpha^{0}(r)=0$, cf. App. 1. \\

{\bf 5. The 2-matrices} \\
The 2-matrix
$\gamma_2(1|1',2|2')=\langle \hat{\psi}^\dag_{1'} \hat{\psi}^\dag_{2'}
\hat{\psi}^{}_2 \hat{\psi}^{}_1 \rangle $
is with
$\hat{N}_{\sigma_2}\hat{\psi_1}=
\hat{\psi}_1(\hat{N}_{\sigma_2}-\delta_{\sigma_1\sigma_2})$, hence
$\hat{N}\hat{\psi_1}=\hat{\psi}_1(\hat{N}-1)$,
normalized as $\text{Tr}\gamma_2=N(N-1)$ with
${\rm Tr}\cdots=\int d1d2\cdots|_{1'=1,2'=2}$. For given positions
$\mathbf{r}^{}_1,\mathbf{r}'^{}_1,\mathbf{r}^{}_2,\mathbf{r}'^{}_2$ there are
16 spin matrix elements. But only six of them are non-zero,
the other 10 matrix elements vanish. With the short-hand notation
$\langle\sigma '_1\sigma '_2\sigma_2\sigma_1\rangle=
\langle\hat{\psi}_{\sigma'_1}^\dag(\mathbf{r}'_1)
\hat{\psi}_{\sigma'_2}^\dag(\mathbf{r}'_2)\hat{\psi}_{\sigma_2}(\mathbf{r}^{}_2)
\hat{\psi}_{\sigma_1}(\mathbf{r}^{}_1)\rangle$
the non-vanishing elements are
\begin{eqnarray}
\gamma_{\uparrow\uparrow} & = & \langle\uparrow\uparrow\uparrow\uparrow\rangle, \;
\quad\qquad {\rm Tr}\gamma_{\uparrow\uparrow}=N_\uparrow(N_\uparrow-1), \nonumber \\
\gamma_{\downarrow\downarrow}&=&\langle\downarrow\downarrow\downarrow\downarrow\rangle, \;
\quad\qquad {\rm Tr}\gamma_{\downarrow\downarrow}=N_\downarrow(N_\downarrow-1), \nonumber \\
\gamma_a^1&=&\langle\uparrow\downarrow\downarrow\uparrow\rangle, \;
\quad\qquad {\rm Tr}\gamma_a^1=N_\downarrow N_\uparrow , \nonumber \\
\gamma_a^2&=&\langle\downarrow\uparrow\uparrow\downarrow\rangle, \;
\quad\qquad {\rm Tr}\gamma_a^2= N_\uparrow N_\downarrow, \nonumber \\
\gamma_a^3&=&\langle\downarrow\uparrow\downarrow\uparrow\rangle, \;
\quad\qquad {\rm Tr}\gamma_a^3= -Na, \nonumber \\
\gamma_a^4&=&\langle\uparrow\downarrow\uparrow\downarrow\rangle, \;
\quad\qquad {\rm Tr}\gamma_a^4= -Na,
\label{6elements}
\end{eqnarray}
whereas e.g. $\langle\uparrow\uparrow\downarrow\downarrow\rangle$,
$\langle\uparrow\uparrow\uparrow\downarrow\rangle$ vanish. Each of these matrix
elements
depends on $\mathbf{r}_1|\mathbf{r}'_1,\mathbf{r}_2|\mathbf{r}'_2$. The first
two lines concern spin-parallel electron pairs and the next four lines describe
spin-antiparallel electron pairs. The properties of these matrix elements
(\ref{6elements}) are listed in Table I together with other related matrix
elements for the spin-antiparallel electrons and with the corresponding
spin-matrices defined below in Eq.(\ref{spinmatrices}). The operators therein
are defined as it follows: D creates
the diagonal elements with $\mathbf{r}'_1=\mathbf{r}_1$ and
$\mathbf{r}'_2=\mathbf{r}_2$, P makes the permutation $\mathbf{r}_1
\leftrightarrow\mathbf{r}_2$, Q effects the hermitian conjugation with
$\mathbf{r}'_1\leftrightarrow\mathbf{r}_1$, $\mathbf{r}'_2\leftrightarrow
\mathbf{r}_2$, and c.c. (= complex conjugation), and C causes the
contraction with $\mathbf{r}'_2\leftrightarrow\mathbf{r}_2$ and $\int d^3r_2$,
where 1-matrices appear, namely $\gamma_\uparrow=\rho_\uparrow f_\uparrow$,
$\gamma_\downarrow=\rho_\downarrow f_\downarrow$, and $\rho\alpha_3$,
$\rho\alpha_4$. For the position matrices $\gamma$, Tr means $\mathbf{r}'_1=
\mathbf{r}_1$, $\mathbf{r}'_2=\mathbf{r}_2$, and $\int d^3r_1 d^3r_2$. For the
spin-matrices $\delta$, Tr means $\sigma'_1=\sigma_1$, $\sigma'_2=\sigma_2$,
and $\sum_{\sigma_{1,2}}$. Note $\gamma_a'=(\gamma_--\gamma_+)/2$ and $\Delta g=
(g_+-g_-)/2$. Columns
for $\tilde {\rm D}$ making $\mathbf{r}'_1=\mathbf{r}_2$,
$\mathbf{r}'_2=\mathbf{r}_1$ and similarly for $\tilde {\rm P}$,
$\tilde {\rm Q}$, $\tilde {\rm C}$
are not shown. - In the column D$\gamma$, i.e.
D$\gamma(\mathbf{r}_1|\mathbf{r}'_1,\mathbf{r}_2|\mathbf{r}'_2)=
\gamma(\mathbf{r}_1|\mathbf{r}_1,\mathbf{r}_2|\mathbf{r}_2)$ the PDs
$g_{\uparrow\uparrow}$, $g_{\downarrow\downarrow}$, $g_a$, and $\Delta g$ are
defined. They depend on $r_{12}=|\mathbf{r}_1-\mathbf{r}_2|$, whereas all the
position matrices depend on $\mathbf{r}_1|\mathbf{r}'_1,\mathbf{r}_2|
\mathbf{r}'_2$. Note D$\gamma_a^1={\rm D}\gamma_a^2$, which holds because of
$\langle{\hat \psi}_\uparrow^\dag(\mathbf{r}_1)
{\hat \psi}_\downarrow^\dag(\mathbf{r}_2)
{\hat \psi}_\downarrow(\mathbf{r}_2)
{\hat \psi}_\uparrow(\mathbf{r}_1)\rangle=
\langle{\hat \psi}_\downarrow^\dag(\mathbf{r}_2)
{\hat \psi}_\uparrow^\dag(\mathbf{r}_1)
{\hat \psi}_\uparrow(\mathbf{r}_1)
{\hat \psi}_\downarrow(\mathbf{r}_2)\rangle$ and $r_{21}=r_{12}$. It is
similarly
D$\gamma_a^3={\rm D}\gamma_a^4$, but for the off-diagonal elements it is
$\gamma_a^1\neq \gamma_a^2$ and $\gamma_a^3\neq \gamma_a^4$. Note further, that
the spin-antiparallel 2-matrices $\gamma_a^1$, $\gamma_a^2$, and $\gamma_a^3$,
$\gamma_a^4$ are indeed hermitian, but have not the Pauli-principle property
P$\gamma=\pm \gamma$. \\

To calculate $V_{ss}$ and $V_a$ we need only the diagonal elements
D$\gamma_{ss}$ and $D\gamma_a^{1,2}$. Also for the contractions only
C$\gamma_{ss}$ and C$\gamma_a^{1,2}$ are needed. So the question arises: what
is the physical
meaning of the spin-exchange matrices $\gamma_a^{3,4}$, which information they
contain ? Because of $\tilde {\rm D} \gamma_a^3=-{\rm D}\gamma_a^1$,
$\tilde {\rm D}\gamma_a^4= -{\rm D} \gamma_a^2$ they contain in principle the same
PD-information as $\gamma_a^{1,2}$.
The 2-matrices $\gamma_a^{3,4}$ mimick spin-flip processes such that
the total spin is conserved. Thus they play a role only if the particle
interaction would be spin-dependent, e.g. different for singlet and triplet
spin-states. The normalization of $\gamma_a^{3,4}$, ${\rm Tr}\gamma_a^{3,4}=
-Na(r_s,\zeta)$, does not appear in the normalization of the total 2-matrix
$\gamma_2$, because of ${\rm Tr}\delta_a^{3,4}=0$ as shown below in
Eq.(\ref{spinmatrices}). For the special cases `no interaction' and `no
polarization' cf. Apps. 1 and 2.  \\

With the spin-matrices (corresponding to the six matrix elements
$\gamma_{ss},\gamma_a^1,\cdots ,\gamma_a^4$, namely)
\begin{eqnarray}
\delta_{\uparrow\uparrow}(\sigma_1|\sigma'_1,\sigma_2|\sigma'_2)&=&
\delta_{\sigma_1,\sigma_1'}\ \delta_{\sigma_2,\sigma_2'}
\delta_{\sigma_1,\uparrow}\delta_{\sigma_2,\uparrow}, \quad \qquad
{\rm Tr}\delta_{\uparrow\uparrow}=1, \nonumber \\
\delta_{\downarrow\downarrow}(\sigma_1|\sigma'_1,\sigma_2|\sigma'_2)&=&
\delta_{\sigma_1,\sigma_1'}\ \delta_{\sigma_2,\sigma_2'}
\delta_{\sigma_1,\downarrow}\delta_{\sigma_2,\downarrow}, \quad\qquad
{\rm Tr}\delta_{\downarrow\downarrow}=1, \nonumber \\
\delta_a^1(\sigma_1|\sigma'_1,\sigma_2|\sigma'_2)&=&
\delta_{\sigma_1,\sigma_1'}\ \delta_{\sigma_2,\sigma_2'}
\delta_{\sigma_1,\uparrow}\delta_{\sigma_2,\downarrow}, \quad\qquad
{\rm Tr}\delta_a^1=1, \nonumber \\
\delta_a^2(\sigma_1|\sigma'_1,\sigma_2|\sigma'_2)&=&
\delta_{\sigma_1,\sigma_1'}\ \delta_{\sigma_2,\sigma_2'}
\delta_{\sigma_1,\downarrow}\delta_{\sigma_2,\uparrow}, \quad\qquad
{\rm Tr}\delta_a^2=1, \nonumber \\
\delta_a^3(\sigma_1|\sigma'_1,\sigma_2|\sigma'_2)&=&
\delta_{\sigma_1,\sigma_2'}\ \delta_{\sigma_2,\sigma_1'}
\delta_{\sigma_1,\uparrow}\delta_{\sigma_2,\downarrow}, \quad\qquad
{\rm Tr}\delta_a^3=0, \nonumber \\
\delta_a^4(\sigma_1|\sigma'_1,\sigma_2|\sigma'_2)&=&
\delta_{\sigma_1,\sigma_2'}\ \delta_{\sigma_2,\sigma_1'}
\delta_{\sigma_1,\downarrow}\delta_{\sigma_2,\uparrow}, \quad\qquad
{\rm Tr}\delta_a^4=0
\label{spinmatrices}
\end{eqnarray}
(note D$\delta_a^{3,4}=0$) the total 2-matrix $\gamma_2$ is given by
\begin{eqnarray}
\gamma_2&=&\delta_{\uparrow\uparrow}\gamma_{\uparrow\uparrow}+
\delta_{\downarrow\downarrow}\gamma_{\downarrow\downarrow}+
\delta_a^1\gamma_a^1+\delta_a^2\gamma_a^2+\delta_a^3\gamma_a^3+
 \delta_a^4\gamma_a^4, \nonumber \\
{\rm Tr}\gamma_2 &=& N_\uparrow (N_\uparrow -1)+N_\downarrow (N_\downarrow -1)
+N_\uparrow N_\downarrow +N_\downarrow N_\uparrow +0+0= N(N-1).
\label{gamma2}
\end{eqnarray}
From Eq.(\ref{gamma2}) also follows the spin-traced 2-matrix
$\gamma_{\uparrow\uparrow}+\gamma_{\downarrow\downarrow}+2\gamma_a$, where
$\gamma_a=\frac{1}{2}(\gamma_a^1+\gamma_a^2)$. \\

In view of the spectral resolution of $\gamma_2$ (respectively of its position
matrices) in terms of antisymmetric spin geminals, the drawback of
Eq.(\ref{gamma2}) is that $\gamma_a^1,\cdots,\gamma_a^4$ are not eigenfunctions
of the permutation operator P as seen from Table I, column P$\gamma$. Only
the linear combinations $\gamma_{\pm}$ have the eigenvalues $\pm$. So it is
useful to rewrite Eq.(\ref{gamma2}) by appropriate (the Pauli symmetry
achieving) linear combinations. In the first (intermediate) step with the
definitions of Table I for $\delta_a, \gamma_a$ etc. (the last but one block of
lines) it results the parallel/antiparallel representation
\begin{eqnarray}
\gamma_2 &=&\delta_{\uparrow\uparrow}\gamma_{\uparrow\uparrow}+
\delta_{\downarrow\downarrow}\gamma_{\downarrow\downarrow}+
\delta_a\gamma_a+\delta'_a\gamma'_a+{\tilde \delta}_a{\tilde \gamma}_a+
 {\tilde \delta}'_a{\tilde \gamma}'_a, \nonumber \\
{\rm Tr}\gamma_2 &=& N_\uparrow (N_\uparrow -1)+N_\downarrow (N_\downarrow -1)
+2N_\uparrow N_\downarrow +0+0+0= N(N-1).
\label{gamma2pa}
\end{eqnarray}
In the next step using $\delta_{\pm},\gamma_\pm$ etc. of Table I (the last block
of lines) it turns out
\begin{eqnarray}
\label{gamma2pm}
\gamma_2 &=&[\delta_{\uparrow\uparrow}\gamma_{\uparrow\uparrow}+
\delta_{\downarrow\downarrow}\gamma_{\downarrow\downarrow}+
\delta_+\gamma_-]+\delta_-\gamma_++[{\tilde \delta}_-{\tilde \gamma}_++
 {\tilde \delta}_+{\tilde \gamma}_-],  \nonumber\\
{\rm Tr}\gamma_2 &=& [N_\uparrow (N_\uparrow -1)+N_\downarrow (N_\downarrow -1)
+(N_\uparrow N_\downarrow -Na)]+(N_\downarrow N_\uparrow+Na)+[0+0]= N(N-1).
\nonumber \\
\end{eqnarray}
This is the triplet/singlet representation of the 2-matrix $\gamma_2$,
because the first three terms describe the spin-triplet, the next term the
spin-singlet, and the last two terms a singlet/triplet mixing, cf. App. 4.
Note, that in the trace the terms $\pm Na$ and in the diagonal elements the
terms $\pm \rho_\uparrow\rho_\downarrow\Delta g$ cancel each other.
The symmetrized, respectively antisymmetrized spin-matrices used in
Eq.(\ref{gamma2pm}) are
\begin{eqnarray}
\delta_\pm(\sigma_1|\sigma'_1,\sigma_2|\sigma'_2)&=&
\frac{1}{2}(\delta_{\sigma_1,\sigma_1'}\ \delta_{\sigma_2,\sigma_2'}\pm
\delta_{\sigma_1,\sigma_2'}\ \delta_{\sigma_2,\sigma_1'})
(\delta_{\sigma_1\uparrow}\delta_{\sigma_2\downarrow}+
\delta_{\sigma_1\downarrow}\delta_{\sigma_2\uparrow}) , \quad
{\rm Tr}\delta_\pm=1,  \nonumber \\
{\tilde \delta}_\pm(\sigma_1|\sigma'_1,\sigma_2|\sigma'_2)&=&
\frac{1}{2}(\delta_{\sigma_1,\sigma_1'}\ \delta_{\sigma_2,\sigma_2'}\mp
\delta_{\sigma_1,\sigma_2'}\ \delta_{\sigma_2,\sigma_1'})
(\delta_{\sigma_1\uparrow}\delta_{\sigma_2\downarrow}-
\delta_{\sigma_1\downarrow}\delta_{\sigma_2\uparrow}), \quad
{\rm Tr}{\tilde \delta}_\pm=0
\label{spinmatrices2}
\end{eqnarray}
and the corresponding position matrices are
\begin{eqnarray}
\gamma_{\pm} &=&\frac{1}{2}(\gamma_a^1+\gamma_a^2\mp\gamma_a^3\mp \gamma_a^4),
\quad {\rm Tr}\gamma_{\pm} =N_\uparrow N_\downarrow \pm Na , \nonumber \\
{\tilde \gamma}_{\pm} &=&
\frac{1}{2}(\gamma_a^1-\gamma_a^2\pm\gamma_a^3\mp \gamma_a^4), \quad
{\rm Tr}{\tilde \gamma}_{\pm} =0,
\label{gammaapm}
\end{eqnarray}
as it follows from Table I.

Hermiticity: Whereas $\gamma_{\uparrow\uparrow}$,
$\gamma_{\downarrow\downarrow}$, and
$\gamma_{\pm}$ are hermitian, the last two terms of Eq.(\ref{gamma2pm}) are
hermitian only summed together because of $Q{\tilde \delta}_\pm=
{\tilde \delta}_\mp$ and $Q{\tilde \gamma}_{\pm}={\tilde \gamma}_{\mp}$. - As
above the spin-traced 2-matrix is again $\gamma_{\uparrow\uparrow}+ \gamma_{\downarrow\downarrow}+2\gamma_a$, because of $\gamma_{+}+\gamma_{-}=\gamma_a^1+
\gamma_a^2=2\gamma_a$. - Note that the interaction energy
$V$ consists of a triplet contribution
$V_{\uparrow\uparrow}+V_{\downarrow\downarrow}+V_{-}$ arising from
$\gamma_{ss}$, $\gamma_{-}$ and of a singlet contribution $V_+$
stemming from $\gamma_{+}$. \\

{\bf 6. The pair densities} \\
For the diagonal elements $\gamma_2(1|1,2|2)=\rho_2(1,2)=\rho^2g(1,2)$,
from Eq.(\ref{gamma2pa}) and Eq.(\ref{gamma2pm}) the two equivalent expressions
($r=r_{12}$)
\begin{eqnarray}
\rho^2g(1,2)&=&\delta_{\sigma_1\uparrow}\delta_{\sigma_2\uparrow}\rho_\uparrow^2
g_{\uparrow\uparrow}(r)+
\delta_{\sigma_1\downarrow}\delta_{\sigma_2\downarrow}\rho_\downarrow^2
g_{\downarrow\downarrow}(r)+
\delta_{\sigma_1,-\sigma_2}
\rho_\uparrow\rho_\downarrow g_a(r) \quad {\rm or} \nonumber \\
\rho^2g(1,2)&=&\left [\delta_{\sigma_1\uparrow}\delta_{\sigma_2\uparrow}\rho_\uparrow^2
g_{\uparrow\uparrow}(r)+
\delta_{\sigma_1\downarrow}\delta_{\sigma_2\downarrow}\rho_\downarrow^2
g_{\downarrow\downarrow}(r)+
\frac{1}{2}\delta_{\sigma_1,-\sigma_2}
\rho_{\uparrow}\rho_{\downarrow}g_-(r)\right ] 
+\frac{1}{2}\delta_{\sigma_1,-\sigma_2}
\rho_{\uparrow}\rho_{\downarrow}g_+(r)\nonumber \\
\label{diag}
\end{eqnarray}
result. The two dimensionless spin-parallel PDs $g_{\uparrow\uparrow}(r),
g_{\downarrow\downarrow}(r)$ are
(with $\mathbf{r}'_1=\mathbf{r}_1$, $\mathbf{r}'_2=\mathbf{r}_2$) defined and
with ${\rm Tr}\gamma_{ss}=N_s(N_s-1)$ normalized by
\begin{eqnarray}
\rho_\uparrow^2g_{\uparrow\uparrow}(r)=
\gamma_{\uparrow\uparrow}(\mathbf{r}^{}_1|\mathbf{r}^{}_1,
\mathbf{r}^{}_2|\mathbf{r}^{}_2), \
\int d^3r\; \rho[1-g_{\uparrow\uparrow}(r)]=\frac{2}{1+\zeta},
\ g_{\uparrow\uparrow}(\infty)=1, \nonumber \\
\rho_\downarrow^2g_{\downarrow\downarrow}(r)=
\gamma_{\downarrow\downarrow}(\mathbf{r}^{}_1|\mathbf{r}^{}_1,
\mathbf{r}^{}_2|\mathbf{r}^{}_2) , \
\int d^3r\; \rho[1-g_{\downarrow\downarrow}(r)]=\frac{2}{1-\zeta},
\ g_{\downarrow\downarrow}(\infty)=1.
\label{paraPD}
\end{eqnarray}
The spin-antiparallel PDs $g_a(r)$, $g_{\pm}(r)$ are defined and
with ${\rm Tr}\gamma_a=N_\uparrow N_\downarrow$,
${\rm Tr}\gamma_{\pm}=N_\uparrow N_\downarrow\pm Na$ normalized as
\begin{eqnarray}
\rho_\uparrow\rho_\downarrow g_a(r)=
\gamma_a(\mathbf{r}_1|\mathbf{r}_1,\mathbf{r}_2|\mathbf{r}_2) , \
\int d^3r\; \rho[1-g_a(r)]=0, \quad g_a(\infty)=1, \nonumber \\
\rho_\uparrow\rho_\downarrow g_{\pm}(r)=
\gamma_{\pm}(\mathbf{r}_1|\mathbf{r}_1,\mathbf{r}_2|\mathbf{r}_2) , \
\int d^3r\; \rho[1-g_{\pm}(r)]=\mp\frac{4a}{1-\zeta^2}, \quad
g_{\pm}(\infty)=1,
\label{antiPD}
\end{eqnarray}
where $4a/(1-\zeta^2)$ equals $2/(1+\zeta)$ for `no interaction' and $2$ for `no
polarization', cf. Apps. 1 and 2. $g_+$ and $g_-$ are the singlet and triplet
spin-antiparallel PDs, respectively. (Here they are normalized as $g_\pm(\infty)
=1$, whereas in Refs. \cite{ZiePR,Ziepss} it is $g_\pm(\infty)=1/2$.) For 
spin-independent particle interaction the PDs
$g_{+}$ and $g_{-}$ are equally weighted, so only their
sum $g_{+}+g_{-}=2g_a$ appears (and not $\frac{1}{2}(g_+-g_-)=\Delta g$, which
is the singlet-triplet splitting of the PDs $g_+$ and $g_-$). Spin summation
yields the spin-weighted PD $g(r)=\sum_{\sigma_{1,2}}g(1,2)$, which is also
given only by $g_{ss}$ and $g_a$:
\begin{eqnarray}
g(r)&=&(\frac{\rho_\uparrow}{\rho})^2g_{\uparrow\uparrow}(r)+
(\frac{\rho_\downarrow}{\rho})^2g_{\downarrow\downarrow}(r)+
2\frac{\rho_\uparrow\rho_\downarrow}{\rho^2} g_a(r) \quad {\rm or} \nonumber \\
1-g(r)&=&(\frac{\rho_\uparrow}{\rho})^2[1-g_{\uparrow\uparrow}(r)]+
(\frac{\rho_\downarrow}{\rho})^2[1-g_{\downarrow\downarrow}(r)]+
2\frac{\rho_\uparrow\rho_\downarrow}{\rho^2}[1-g_a(r)]
\label{gofr}
\end{eqnarray}
with $\rho_\uparrow/\rho=(1+\zeta)/2$ and $\rho_\downarrow/\rho=(1-\zeta)/2$ and
$\int d^3r\; \rho[1-g(r)]=1, g(\infty)=1$. \\

`On-top', i.e. for $r=0$, the coalescing cusp theorems
\begin{equation}
g'''_{ss}(0)=\frac{3}{2}\alpha_0 g''_{ss}(0), \quad
g'_a(0)=\alpha_0 g_a(0), \quad
\alpha_0= \left (\frac{4}{9\pi}\right )^{1/3}r_s
\end{equation}
hold \cite{Kim1,Raja}. \\

{\bf 7. The contraction SRs} \\
The contraction of the spin-parallel 2-matrices,
\begin{eqnarray}
&&\int d^3r_2\gamma_{ss}
(\mathbf{r}^{}_1|\mathbf{r}'^{}_1,\mathbf{r}^{}_2|\mathbf{r}^{}_2)=
\gamma_{s}(\mathbf{r}^{}_1|\mathbf{r}'^{}_1)
(N_{s}-1)\quad{\rm or} \ \nonumber \\
&&\int d^3r_2[\rho_{s} \gamma_{s}(\mathbf{r}_1|\mathbf{r}'^{}_1)
-\gamma_{ss}(\mathbf{r}^{}_1|\mathbf{r}'^{}_1,\mathbf{r}^{}_2|
\mathbf{r}^{}_2)]=\gamma_s(\mathbf{r}_1|\mathbf{r}'^{}_1),
\label{contractionpara}
\end{eqnarray}
makes the spin-parallel 3-point functions $\gamma_{ss}
(\mathbf{r}^{}_1|\mathbf{r}'_1, \mathbf{r}^{}_2|\mathbf{r}^{}_2)$ to appear.
Because of translational and rotational invariance (six conditions) this
function
depends only on the three distances $r_{11'}=|\mathbf{r}_1-\mathbf{r}'_1|$,
$r_{12}=|\mathbf{r}_1-\mathbf{r}_2|$, and
$r_{1'2}=|\mathbf{r}'_1-\mathbf{r}_2|$, besides the direction of
$\mathbf{r}_{12}$ can be averaged:
\begin{equation}
G_{ss}(r_{11'},r_{12})=\int\frac{d\Omega_{12}}{4\pi}
G_{ss}(r_{11'},r_{12},|\mathbf{r}_{11'}-\mathbf{r}_{12}|), \quad
\rho_{s}^2G_{ss}(r_{11'},r_{12},r_{1'2})=
\gamma_{ss}(\mathbf{r}_1|\mathbf{r}'_1,
\mathbf{r}^{}_2|\mathbf{r}^{}_2) .
\label{paraG}
\end{equation}
$G_{\uparrow\uparrow}(r_{11'},r_{12})$ and
$G_{\downarrow\downarrow}(r_{11'},r_{12})$ are
special 2-matrices yielding with $r'_1=r_1$ the spin-parallel PDs:
$g_{ss}(r_{12})=G_{ss}(0,r_{12})$.
The contraction SRs (\ref{contractionpara}) in terms of these 2-matrices
$G_{ss}(r_{11'},r_{12})$ are rewritten as
\begin{equation}
\int d^3r_{12}\; \rho_s[f_s(r_{11'})-
G_{ss}(r_{11'},r_{12})]=
f_{s}(r_{11'}), \quad
G_{ss}(r_{11'},\infty)=f_{s}(r_{11'}).
\label{paracontraction}
\end{equation}
So the functions $G_{ss}$ contain not only $g_{ss}(r)$, but
also $f_s(r)$ and with $r_{11'}=0$ the SRs (\ref{paracontraction}) yield
the normalization (\ref{paraPD}) of $g_{ss}(r)$. \\

The contraction of the spin-antiparallel 2-matrices $\gamma_{\pm}$ yields
\begin{eqnarray}
&&\int d^3r_2\gamma_{\pm}(\mathbf{r}_1|\mathbf{r}'_1,\mathbf{r}_2|\mathbf{r}_2)=
\frac{1}{2}[N_\downarrow\gamma_\uparrow(\mathbf{r}_1|\mathbf{r}'_1)+
N_\uparrow\gamma_\downarrow(\mathbf{r}_1|\mathbf{r}'_1)]\pm\rho\alpha(r_{11'})\quad {\rm or},
\nonumber \\
&&\int d^3r_2 \left \{\frac{1}{2}[\rho_\downarrow
\gamma_\uparrow+\rho_\uparrow\gamma_\downarrow](\mathbf{r}_1|\mathbf{r}'_1)
-\gamma_a(\mathbf{r}_1|\mathbf{r}'_1,\mathbf{r}_2|\mathbf{r}_2) \right \}=
\mp\rho\alpha(r_{11'}).
\label{contractionanti}
\end{eqnarray}
Corresponding 2-point functions are defined by
\begin{equation}
\rho_\uparrow \rho_\downarrow G_{\pm}(r_{11'},r_{12})=
\int\frac{d\Omega_{12}}{4\pi}
\gamma_{\pm}(\mathbf{r}_1|\mathbf{r}'_1,\mathbf{r}_2|\mathbf{r}_2).
\label{Ga}
\end{equation}
With this definition of $G_{\pm}$ the contraction SRs
(\ref{contractionanti}) can be rewritten as
\begin{equation}
\int d^3r_{12}\; \rho [f(r_{11'})-G_{\pm}(r_{11'},r_{12})]=
\mp\frac{4\alpha(r_{11'})}{1-\zeta^2}, \quad
G_{\pm}(r_{11'},\infty)=f(r_{11'})
\label{anticontraction1}
\end{equation}
with $f(r)=\frac{1}{2}[f_\uparrow(r)+f_\downarrow(r)]$.
The functions $G_{\pm}$ contain the spin-antiparallel PDs $g_{\pm}(r_{12})=
G_{\pm}(0,r_{12})$. Thus the SRs (\ref{anticontraction1})
contain with $r_{11'}=0$ the normalizations
(\ref{antiPD}) of $g_{\pm}(r)$. Note that the 1-matrix $\alpha(r_{11'})$
determines the normalization of the PDs $g_{\pm}$ with $\alpha(0)=a$.
- For $G_{a}=(G_{+}+G_{-})/2$, which equally
weights the singlet/triplet contributions, it holds
\begin{equation}
\int d^3r_{12}\; \rho[f(r_{11'})- G_{a}(r_{11'},r_{12})]=0, \quad
G_{a}(r_{11'},\infty)=f(r_{11'}).
\label{anticontraction2}
\end{equation}
In summary, the two functions $G_{ss}(r_{11'},r_{12})$ not only
contain the
two spin-parallel PDs $g_{ss}(r)=G_{ss}(0,r)$, but also the
two 1-matrices $f_s(r)=G_{ss}(r,\infty)$. And the function
$G_a(r_{11'},r_{12})$ contains the spin-antiparallel PD
$g_a(r)=G_a(0,r)$ and besides it controls $f=
\frac{1}{2}[f_\uparrow+f_\downarrow]$ with $\frac{1}{2}[f_\uparrow(r)+
f_\downarrow(r)]= G_a(r,\infty)$. A more refined description distinguishes
between the singlet function $G_{+}$ and the triplet function $G_{-}$. So
there are six SRs, namely
the four contraction SRs (\ref{paracontraction}) and (\ref{anticontraction1})
and the two momentum-distribution normalizations $f_s(0)=1$. \\

{\bf 8. Wick's theorem, linked diagrams, and the cumulant expansion} \\
If the Coulomb repulsion is treated as perturbation (with many-body $S$-matrix
theory and Wick's theorem), then the 1-matrices $\gamma_\uparrow$ and
$\gamma_\downarrow$ are given by $linked$ Feynman diagrams only, as sketched in
(Eq.(\ref{onematrix}) shows with $\gamma_1\sim\delta_{\sigma_1,\sigma_2}$:
there is no spin-flip along an electron line)

\begin{fmffile}{1stgraph}
\begin{eqnarray}
\langle \hat{\psi}^{\dag}_s(\mathbf{r}'_1)\hat{\psi}_s(\mathbf{r}_1)
\rangle = \quad
\parbox{20mm}{
   \fmfframe(1,2)(1,2){
   \begin{fmfgraph*}(20,12)
   \fmfipair{am,bm,cm,dm,em,fm}
   \fmfiequ{am}{sw}
   \fmfiequ{bm}{se}
   \fmfiequ{cm}{ne}
   \fmfiequ{dm}{nw}
   \fmfiequ{em}{0.5[nw,sw]}
   \fmfiequ{fm}{0.5[ne,se]}
   \fmfi{fermion,lab=$s$,lab.sid=left}{fm--em}
   \fmfiv{d.siz=3thin,lab=$1'$}{fm}
   \fmfiv{d.siz=3thin,lab=$1$}{em}
   \fmfiv{d.sh=circle,d.siz=3thin}{fm}
   \fmfiv{d.sh=circle,d.siz=3thin}{em}
   \end{fmfgraph*}}}
   \quad \quad + \quad
\parbox{20mm}{
   \fmfframe(1,2)(1,2){
   \begin{fmfgraph*}(20,12)
   \fmfipair{am,bm,cm,dm,em,fm,gm,hm,im,jm}
   \fmfiequ{am}{sw}
   \fmfiequ{bm}{se}
   \fmfiequ{cm}{ne}
   \fmfiequ{dm}{nw}
   \fmfiequ{em}{0.5[nw,sw]}
   \fmfiequ{fm}{0.5[ne,se]}
   \fmfiequ{gm}{1/3[em,fm]} \fmfiequ{hm}{2/3[em,fm]}
   \fmfiequ{im}{0.5[nw,ne]}
   \fmfiequ{jm}{im-(0,.1h)}
   \fmfi{fermion,lab=$s$,lab.sid=left}{fm--hm}
   \fmfi{fermion,lab=$s$,lab.sid=left}{hm--gm}
   \fmfi{fermion,lab=$s$,lab.sid=left}{gm--em}
   \fmfi{dashes}{hm{fm-hm}..tension 2..{left}jm}
   \fmfi{dashes}{jm{left}..tension 2..{gm-em}gm}
   \fmfiv{d.siz=3thin,lab=$1'$}{fm}
   \fmfiv{d.siz=3thin,lab=$1$}{em}
   \fmfiv{d.sh=circle,d.siz=3thin}{fm}
   \fmfiv{d.sh=circle,d.siz=3thin}{em}
   \fmfiv{d.sh=circle,d.siz=3thin}{gm}
   \fmfiv{d.sh=circle,d.siz=3thin}{hm}
   \end{fmfgraph*}}} \quad \quad + \cdots \; .
\end{eqnarray}
\end{fmffile}

\noindent
Contrary to this the 2-matrices are given by
$both$ unlinked (or disconnected) $and$ linked (or connected) diagrams. If all
the unlinked diagrams are summed up, products of the 1-matrices appear, e.g.
\begin{eqnarray}
\gamma_{ss}=\langle ssss\rangle=
 \gamma_s(\mathbf{r}_1|\mathbf{r}'_1)\gamma_s(\mathbf{r}_2|\mathbf{r}'_2)
 -\gamma_s(\mathbf{r}_1|\mathbf{r}'_2)\gamma_s(\mathbf{r}_2|\mathbf{r}'_1)+
\langle ssss \rangle_{\rm c},
\label{cum1}
\end{eqnarray}
where $\langle ssss \rangle_c=-\chi_{ss}$ is given by the sum of all the
connected diagrams of

\begin{fmffile}{2ndgraph}
\begin{eqnarray}
\chi_{ss}=\langle \hat{\psi}^{\dag}_s(\mathbf{r}'_1)\hat{\psi}^{\dag}_s(\mathbf{r}'_2)
\hat{\psi}_s(\mathbf{r}_2)\hat{\psi}_s(\mathbf{r}_1)
\rangle_c = \quad
\parbox{20mm}{
   \fmfframe(1,2)(1,2){
   \begin{fmfgraph*}(20,12)
   \fmfipair{am,bm,cm,dm,em,fm}
   \fmfiequ{am}{sw}
   \fmfiequ{bm}{se}
   \fmfiequ{cm}{ne}
   \fmfiequ{dm}{nw}
   \fmfiequ{em}{0.5[nw,ne]}
   \fmfiequ{fm}{0.5[sw,se]}
   \fmfi{fermion,lab=$s$,lab.sid=right}{cm--em}
   \fmfi{fermion,lab=$s$,lab.sid=right}{em--dm}
   \fmfi{fermion,lab=$s$,lab.sid=left}{bm--fm}
   \fmfi{fermion,lab=$s$,lab.sid=left}{fm--am}
   \fmfi{dashes}{em--fm}
   \fmfiv{d.siz=3thin,lab=$1'$}{cm}
   \fmfiv{d.siz=3thin,lab=$1$}{dm}
   \fmfiv{d.siz=3thin,lab=$2'$}{bm}
   \fmfiv{d.siz=3thin,lab=$2$}{am}
   \fmfiv{d.sh=circle,d.siz=3thin}{cm}
   \fmfiv{d.sh=circle,d.siz=3thin}{dm}
   \fmfiv{d.sh=circle,d.siz=3thin}{bm}
   \fmfiv{d.sh=circle,d.siz=3thin}{am}
   \fmfiv{d.sh=circle,d.siz=3thin}{em}
   \fmfiv{d.sh=circle,d.siz=3thin}{fm}
   \end{fmfgraph*}}}
   \quad + \quad
\parbox{20mm}{
   \fmfframe(1,2)(1,2){
   \begin{fmfgraph*}(20,12)
   \fmfipair{am,bm,cm,dm,em,fm}
   \fmfiequ{am}{sw}
   \fmfiequ{bm}{se}
   \fmfiequ{cm}{ne}
   \fmfiequ{dm}{nw}
   \fmfiequ{em}{0.5[nw,ne]}
   \fmfiequ{fm}{0.5[sw,se]}
   \fmfi{fermion,lab=$s$,lab.sid=right}{cm--em}
   \fmfi{fermion,lab=$s$,lab.sid=right}{em--dm}
   \fmfi{fermion,lab=$s$,lab.sid=left}{bm--fm}
   \fmfi{fermion,lab=$s$,lab.sid=left}{fm--am}
   \fmfi{dashes}{em--fm}
   \fmfiv{d.siz=3thin,lab=$1'$}{bm}
   \fmfiv{d.siz=3thin,lab=$1$}{dm}
   \fmfiv{d.siz=3thin,lab=$2'$}{cm}
   \fmfiv{d.siz=3thin,lab=$2$}{am}
   \fmfiv{d.sh=circle,d.siz=3thin}{cm}
   \fmfiv{d.sh=circle,d.siz=3thin}{dm}
   \fmfiv{d.sh=circle,d.siz=3thin}{bm}
   \fmfiv{d.sh=circle,d.siz=3thin}{am}
   \fmfiv{d.sh=circle,d.siz=3thin}{em}
   \fmfiv{d.sh=circle,d.siz=3thin}{fm}
   \end{fmfgraph*}}} \quad \quad + \cdots \; .
\label{diagr2}
\end{eqnarray}
\end{fmffile}

\noindent
The index `c' means `connected' or `cumulant'. It is similarly
\begin{eqnarray}
\gamma_a^1=\langle\uparrow\downarrow\downarrow\uparrow\rangle=
\gamma_\uparrow(\mathbf{r}_1|\mathbf{r}'_1)\gamma_\downarrow
(\mathbf{r}_2|\mathbf{r}'_2)+
\langle\uparrow\downarrow\downarrow\uparrow\rangle_{\rm c}, \nonumber \\
\gamma_a^3=\langle\downarrow\uparrow\downarrow\uparrow\rangle=
\gamma_\uparrow(\mathbf{r}_1|\mathbf{r}'_2)\gamma_\downarrow
(\mathbf{r}_2|\mathbf{r}'_1)+
\langle\downarrow\uparrow\downarrow\uparrow\rangle_{\rm c},
\label{cum2}
\end{eqnarray}
where $\langle\uparrow\downarrow\downarrow\uparrow\rangle_{\rm c}=-\chi_a^1$
and $\langle\downarrow\uparrow\downarrow\uparrow\rangle_{\rm c}=-\chi_a^3$
mean the linked diagrams of

\begin{fmffile}{3rdgraph}
\begin{eqnarray}
\chi^1_a=\langle \hat{\psi}^{\dag}_{\uparrow}(\mathbf{r}'_1)\hat{\psi}^{\dag}_{\downarrow}(\mathbf{r}'_2)
\hat{\psi}_{\downarrow}(\mathbf{r}_2)\hat{\psi}_{\uparrow}(\mathbf{r}_1)
\rangle_c =& \quad
\parbox{20mm}{
   \fmfframe(1,2)(1,2){
   \begin{fmfgraph*}(20,12)
   \fmfipair{am,bm,cm,dm,em,fm}
   \fmfiequ{am}{sw}
   \fmfiequ{bm}{se}
   \fmfiequ{cm}{ne}
   \fmfiequ{dm}{nw}
   \fmfiequ{em}{0.5[nw,ne]}
   \fmfiequ{fm}{0.5[sw,se]}
   \fmfi{fermion,lab=$\uparrow$,lab.sid=right}{cm--em}
   \fmfi{fermion,lab=$\uparrow$,lab.sid=right}{em--dm}
   \fmfi{fermion,lab=$\downarrow$,lab.sid=left}{bm--fm}
   \fmfi{fermion,lab=$\downarrow$,lab.sid=left}{fm--am}
   \fmfi{dashes}{em--fm}
   \fmfiv{d.siz=3thin,lab=$1'$}{cm}
   \fmfiv{d.siz=3thin,lab=$1$}{dm}
   \fmfiv{d.siz=3thin,lab=$2'$}{bm}
   \fmfiv{d.siz=3thin,lab=$2$}{am}
   \fmfiv{d.sh=circle,d.siz=3thin}{cm}
   \fmfiv{d.sh=circle,d.siz=3thin}{dm}
   \fmfiv{d.sh=circle,d.siz=3thin}{bm}
   \fmfiv{d.sh=circle,d.siz=3thin}{am}
   \fmfiv{d.sh=circle,d.siz=3thin}{em}
   \fmfiv{d.sh=circle,d.siz=3thin}{fm}
   \end{fmfgraph*}}}  \quad +& \cdots \; ,\nonumber \\ \nonumber\\ \nonumber\\
\chi^3_a=\langle \hat{\psi}^{\dag}_{\downarrow}(\mathbf{r}'_1)\hat{\psi}^{\dag}_{\uparrow}(\mathbf{r}'_2)
\hat{\psi}_{\downarrow}(\mathbf{r}_2)\hat{\psi}_{\uparrow}(\mathbf{r}_1)
\rangle_c =& \quad
\parbox{20mm}{
   \fmfframe(1,2)(1,2){
   \begin{fmfgraph*}(20,12)
   \fmfipair{am,bm,cm,dm,em,fm}
   \fmfiequ{am}{sw}
   \fmfiequ{bm}{se}
   \fmfiequ{cm}{ne}
   \fmfiequ{dm}{nw}
   \fmfiequ{em}{0.5[nw,ne]}
   \fmfiequ{fm}{0.5[sw,se]}
   \fmfi{fermion,lab=$\uparrow$,lab.sid=right}{cm--em}
   \fmfi{fermion,lab=$\uparrow$,lab.sid=right}{em--dm}
   \fmfi{fermion,lab=$\downarrow$,lab.sid=left}{bm--fm}
   \fmfi{fermion,lab=$\downarrow$,lab.sid=left}{fm--am}
   \fmfi{dashes}{em--fm}
   \fmfiv{d.siz=3thin,lab=$1'$}{bm}
   \fmfiv{d.siz=3thin,lab=$1$}{dm}
   \fmfiv{d.siz=3thin,lab=$2'$}{cm}
   \fmfiv{d.siz=3thin,lab=$2$}{am}
   \fmfiv{d.sh=circle,d.siz=3thin}{cm}
   \fmfiv{d.sh=circle,d.siz=3thin}{dm}
   \fmfiv{d.sh=circle,d.siz=3thin}{bm}
   \fmfiv{d.sh=circle,d.siz=3thin}{am}
   \fmfiv{d.sh=circle,d.siz=3thin}{em}
   \fmfiv{d.sh=circle,d.siz=3thin}{fm}
   \end{fmfgraph*}}} \quad +& \cdots \; .
\label{diagr3}
\end{eqnarray}
\end{fmffile}

\noindent
The products of the 1-matrices $\gamma_s=\rho_s f_s$ may be called generalized
Hartree-Fock (HF) parts, where the term `generalized' means `built up from the
true, hence non-idempotent 1-matrix $f_s(r)$'. The HF parts of the six position
matrices (\ref{6elements}) together with eight other related 2-matrices and
their properties are listed in Table II. The matrices $\chi_{ss}$, $\chi_a^1$,
$\chi_a^3$ etc. are called cumulant 2-matrices, they are listed in Table III.
Eqs.(\ref{cum1}) and (\ref{cum2}) are the first steps of the more general
cumulant expansion \cite{foo2}. Per construction, the cumulant matrices
are given by linked diagrams, what
makes them size-extensively normalizable. Contrary to this
the normalizations of the $\gamma$-matrices and the $\gamma^{\rm HF}$-matrices
contain terms $\sim N^2$, cf. the columns `Tr' of Table III with those of
Tables I and II. In the column C$\gamma^{\rm HF}$ of Table II the already above
introduced 1-matrices $\beta_{s}(r_{11'})$ and $\beta_a(r_{11'})$
appear, as it is obvious from Eq.(\ref{beta}). From the diagonal elements
D$\gamma^{\rm HF}$ follow the PDs $g_{ss}^{\rm HF}(r)=
1-|f_s(r)|^2$ with $\rho_s\int d^3r [1-g_{ss}^{\rm HF}(r)]=\frac{N}{N_s}b_{s}$
and $g_a^{\rm HF}(r)=1$ as well as ${\Delta g}^{\rm HF}(r)=
f_\uparrow(r) f_\downarrow(r)$. \\

The diagonal elements of $\chi_{ss}$ and $\chi_a^1$ or $\chi_a^2$ give the
cumulant contributions to the PDs:
\begin{eqnarray}
\rho_{s}^2h_{ss}& = & {\rm D}\chi_{s},\;
\int d^3r\; \rho h_{ss}(r)= \frac{\rho}{\rho_s}(1-\frac{\rho}{\rho_s}b_{s}), \;
g_{ss}(r) =1 -  |f_s(r)|^2  -h_{s}(r),
\nonumber \\
\rho_\uparrow\rho_\downarrow h_{\pm} & =& {\rm D}\chi_{\pm},
\int d^3r\; \rho h_{\pm}(r)=\mp (a-b_a),
g_{\pm}(r)=1\mp [\alpha(r)-\beta_a(r)]-h_{\pm}(r),
\nonumber \\
\rho_\uparrow\rho_\downarrow h_a & = & {\rm D} \chi_a,\;
\int d^3r\; \rho h_a(r)=0, \;
g_a(r) =1-  0  -h_a(r).
\end{eqnarray}
The contraction of the cumulant 2-matrix: With the definitions
\begin{eqnarray}
\rho_s^2H_{ss}(r_{11'},r_{12})&=&
\int\frac{d\Omega_{12}}{4\pi}\int d^3r_2
\chi_{ss}(\mathbf{r}_1|\mathbf{r}'_1,\mathbf{r}_2|\mathbf{r}_2), \nonumber \\
\rho_\uparrow\rho_\downarrow H_{\pm}(r_{11'},r_{12})&=&
\int\frac{d\Omega_{12}}{4\pi}\int d^3r_2
\chi_{\pm}(\mathbf{r}_1|\mathbf{r}'_1,\mathbf{r}_2|\mathbf{r}_2),
\label{cumucon}
\end{eqnarray}
and $H_a=(H_{+}+H_{-})/2$ the contraction SRs are
\begin{eqnarray}
\int d^3r_{12}\; \rho H_{ss}(r_{11'},r_{12}) & = &
\frac{\rho}{\rho_s}[f_s(r_{11'})-\frac{\rho}{\rho_s}\beta_{s}(r_{11'})],
\nonumber \\
\int d^3r_{12}\; \rho H_{\pm}(r_{11'},r_{12}) & = & \mp
\frac{\rho^2}{\rho_\uparrow\rho_\downarrow}
[\alpha(r_{11'})-\beta_a(r_{11'})], \nonumber \\
\int d^3r_{12}\;\rho H_a(r_{11'},r_{12}) & = & 0.
\end{eqnarray}
The advantage of the $H$-functions: in perturbation theory they are given by
linked diagrams, which are size-extensive. They contain the cumulant PDs
$h_{ss}(r)=H_{ss}(0,r)$,
$h_{\pm}(r)=H_{\pm}(0,r)$, $h_{a}(r)=H_{a}(0,r)$. To get also the 1-matrices
$f_s$ (or equivalently the momentum distributions $n_s$) one has to write the
first equation of the SRs (\ref{cumucon}) as
\begin{equation}
\int d^3r_{12}\; \rho H_{ss}(r_{11'},r_{12})=
\left (\frac{\rho}{\rho_s}\right )^2
\frac{1}{N}\sum_\mathbf{k}n_s(k)[1-n_s(k)]{\rm e}^{i\mathbf{k}\mathbf{r}_{11'}}
\end{equation}
and to Fourier analyse the lhs according to
\begin{equation}
\int d^3r_{12}\; \rho H_{ss}(r_{11'},r_{12})=
\left ( \frac{\rho}{\rho_s}\right )^2\frac{1}{N}\sum_\mathbf{k}
m_s(k){\rm e}^{i\mathbf{k}\mathbf{r}_{11'}}
\end{equation}
defining functions $m_s(k)$. Then one has to solve the quadratic equation
$n_s^2(k)-n_s(k)+m_s(k)=0$. As already mentioned, one way to get the PDs and
the momentum distributions is to perturbatively calculate linked diagrams for
$H_{ss}$ and $H_{\pm}$. Another way is to solve an effective 2-body
Schr\"odinger equation for the geminals, which parametrize the 2-matrices
$\gamma_{ss}$ and $\gamma_{\pm}$ like this has been done successfully for the
spin-unpolarized HEG \cite{Kim1,Over,Gor3,Dav}. This way is sketched in the
following for the spin-polarized HEG. \\

{\bf 9. The spectral resolutions} \\
The four 2-matrices $\gamma_{ss}$ and $\gamma_{\pm}$ are needed for the key
functions $G_{ss}$ and $G_{\pm}$. They have the correct Pauli
symmetry and are hermitian. So they can be diagonalized in terms of
symmetric/antisymmetric geminals and corresponding occupancies. This is similar
to the well-known natural representation of the 1-matrix. For the HF
parts this diagonalization can be done easily with the help of the spectral
resolution
or natural-orbital representation (\ref{1matrix}) of the 1-matrices $f_s(r)$ in
terms of plane waves (because the system is homogeneous). Thereby the geminals
\begin{equation}
\frac{1}{\sqrt \Omega}{\rm e}^{{\rm i}\mathbf{K}\mathbf{R}}
\frac{1}{\sqrt \Omega}{\rm e}^{{\rm i}\mathbf{k}\mathbf{r}} \quad {\rm with} \quad
\mathbf{R}=\frac{1}{2}(\mathbf{r}_1+\mathbf{r}_2), \;
\mathbf{r}=\mathbf{r}_1-\mathbf{r}_2
\end{equation}
appear together with the weights
\begin{eqnarray}
\mu_{s}(\mathbf{K},\mathbf{k})=n_s(k_1)n_{s}(k_2)&,& \quad
\mu_{a}(\mathbf{K},\mathbf{k})= n_\uparrow(k_1)n_\downarrow(k_2) \nonumber \\
{\rm with} \quad
\mathbf{k}_1=\frac{1}{2}\mathbf{K}+\mathbf{k}&,& \;
\mathbf{k}_2=\frac{1}{2}\mathbf{K}-\mathbf{k}.
\label{weights}
\end{eqnarray}
$\mathbf{R}$ is the centre-of-mass coordinate and $\frac{1}{\sqrt \Omega}{\rm e}^{{\rm i}\mathbf{K}\mathbf{R}}$ describes the centre-of-mass motion (which is a
free-particle motion, because the system is homogeneous), whereas $\mathbf{r}$
is the relative coordinate and $\varphi^0(\mathbf{r},\mathbf{k})=
\frac{1}{\sqrt \Omega}{\rm e}^{{\rm i}\mathbf{k}\mathbf{r}}$ describes the
relative motion of the HF parts. Starting with the expressions for
$\gamma_{ss}^{\rm HF}$ and $\gamma_{\pm}^{\rm HF}$ of Table II, the following
spectral resolutions of the HF parts result
\begin{eqnarray}
\gamma_{ss}^{\rm HF}(\mathbf{R}|\mathbf{R}',\mathbf{r}|\mathbf{r}')=
\sum_{\mathbf{K},\mathbf{k}}\mu_{s}(\mathbf{K},\mathbf{k})
\varphi_-^0(\mathbf{R},\mathbf{r})
\varphi_-^{0*}(\mathbf{R}',\mathbf{r}') ,
\nonumber \\
\gamma_{\pm}^{\rm HF}(\mathbf{R}|\mathbf{R}',\mathbf{r}|\mathbf{r}')=
\sum_{\mathbf{K},\mathbf{k}}\mu_{a}(\mathbf{K},\mathbf{k})
\varphi_\pm^0(\mathbf{R},\mathbf{r})
\varphi_\pm^{0*}(\mathbf{R}',\mathbf{r}')
\label{specres}
\end{eqnarray}
with $\varphi_\pm^0(\mathbf{R},\mathbf{r})=
\frac{1}{\sqrt \Omega}{\rm e}^{{\rm i}\mathbf{K}\mathbf{R}}
\frac{4\pi}{\sqrt\Omega}
\sum_L^{\pm}{\rm i}^lj_L(k\mathbf{r})Y_L^*(\mathbf{e}_k)$, where the dependence
of $\mathbf{K},\mathbf{k}$ is not explicitly denoted and the angular
momentum expansion of a plane wave is used. Note $L=(l,m_l)$ and
$j_L(k\mathbf{r})=j_l(kr)Y_L(\mathbf{e}_r)$. So the spectral resolution
(\ref{specres}) can be written in terms of free-electron geminals $j_l(kr)$ and
occupancy matrices $\mu_{LL'}^{s}(\mathbf{K},k)$ and
$\mu_{LL'}^{a}(\mathbf{K},k)$.
The normalizations
\begin{eqnarray}
{\rm Tr}\gamma_{ss}^{\rm HF} &=&
\sum_{\mathbf{k}_{1,2}}\mu_{s}(\mathbf{K},\mathbf{k})
[1-\delta_{\mathbf{k}_1,\mathbf{k}_2}]=
N_s^2-\sum_\mathbf{k}n_s^2(k)=N_s^2-Nb_{s},
\nonumber \\
{\rm Tr}\gamma_{\pm}^{\rm HF} &=& \sum_{\mathbf{k}_{1,2}}
\mu_a(\mathbf{K},\mathbf{k})[1\pm\delta_{\mathbf{k}_1,\mathbf{k}_2}]=
N_\uparrow N_\downarrow\pm\sum_\mathbf{k}n_\uparrow(k)n_\downarrow(k)=
N_\uparrow N_\downarrow\pm Nb_a
\end{eqnarray}
follow from $\int\frac{d^3r}{\Omega}{\rm e}^{{\rm i}\mathbf{k}\mathbf{r}}=
\delta_{\mathbf{k},0}$ in agreement with Table II.

From the spectral resolution (\ref{specres}) of the HF parts one may derive
the spectral resolution of the total 2-matrix by replacing the
the free-electron geminals $\varphi_\pm^0$ or $j_l(kr)$ by interacting-electron
geminals $\varphi_\pm$ or $R_l(r,k)$:
\begin{eqnarray}
\gamma_{ss}(\mathbf{R}|\mathbf{R}',\mathbf{r}|\mathbf{r}')=
\sum_{\mathbf{K},\mathbf{k}}\mu_{s}(\mathbf{K},\mathbf{k})
\varphi_-(\mathbf{R},\mathbf{r})
\varphi_-^{*}(\mathbf{R}',\mathbf{r}') ,
\nonumber \\
\gamma_{\pm}(\mathbf{R}|\mathbf{R}',\mathbf{r}|\mathbf{r}')=
\sum_{\mathbf{K},\mathbf{k}}\mu_{a}(\mathbf{K},\mathbf{k})
\varphi_\pm(\mathbf{R},\mathbf{r})
\varphi_\pm^{*}(\mathbf{R}',\mathbf{r}').
\label{specres2}
\end{eqnarray}
(The general structure (\ref{specres2}) in terms of eigenfunctions and
eigenvalues, which diagonalize $\gamma_{ss}$ and $\gamma_{\pm}$, is correct, 
but the use of Eq.(\ref{weights}) for the geminal weights, which is right in 
Eq.(\ref{specres}), may be here in Eq.(\ref{specres2}) only an approximation.) 
With the angular-momentum expansion of $\varphi_\pm(\mathbf{R},
\mathbf{r})$ the PDs $g_{ss}$ and $g_{\pm}$ are given in terms of the afore mentioned 
geminals $R_l(r,k)$ and occupancies $\mu_{LL'}^{s}({\mathbf{K},k})$
and $\mu_{LL'}^{a}({\mathbf{K},k})$. Their diagonalization causes a
$(\mathbf{K},k)$-dependent $L$-mixing with $L$-mixed geminals
$R_\Lambda^\pm(\mathbf{r},k)$ and occupancies $\mu_\Lambda^\pm$. Calculating
the PDs with the non-spherically
symmetric functions $|R_\Lambda^\pm(\mathbf{r},k)|^2$, their non-sphericity
is averaged out by summing over $\Lambda$ with $\mu_\Lambda^\pm$.
The geminals $R_l(r,k)$ are the scattering state solutions
of a 2-body Schr\"odinger equation with an effective
repulsion $v_\pm(r)=\frac{\epsilon^2}{r}+\cdots$, which may be in general a
non-local interaction potential, but a local one possibly  can be a good
approximation. Whithin a PDFT the ellipsis result from a functional derivative
of the kinetic energy as a functional the PD \cite{Zie4}. \\

Eq.(\ref{specres2}) means for the cumulant matrices
\begin{eqnarray}
\chi_{ss}(\mathbf{R}|\mathbf{R}',\mathbf{r}|\mathbf{r}')=
\sum_{\mathbf{K},\mathbf{k}}\mu_{s}(\mathbf{K},\mathbf{k})
[\varphi_-^0(\mathbf{R},\mathbf{r}){\varphi_-^{0*}}(\mathbf{R}',\mathbf{r}')-
\varphi_-(\mathbf{R},\mathbf{r}){\varphi_-^*}(\mathbf{R}',\mathbf{r}')] ,
\nonumber \\
\chi_{\pm}(\mathbf{R}|\mathbf{R}',\mathbf{r}|\mathbf{r}')=
\sum_{\mathbf{K},\mathbf{k}}\mu_{a}(\mathbf{K},\mathbf{k})
[\varphi_\pm^0(\mathbf{R},\mathbf{r}){\varphi_\pm^{0*}}(\mathbf{R}',\mathbf{r}')-
\varphi_\pm(\mathbf{R},\mathbf{r}){\varphi_\pm^*}(\mathbf{R}',\mathbf{r}')].
\end{eqnarray}
Because the cumulant matrices $\chi_{ss}$ and $\chi_{\pm}$ are hermitian,
they can be represented alternatively in their `own' spectral resolution
with cumulant weights $\mu_{s,a}^c$ and cumulant geminals $R_l^c$ being the
bound state solutions of another
2-body Schr\"odinger equation with an effective attraction $v_\pm^{\rm c}(r)$,
possibly $v_\pm^{\rm c}(r)=-v_\pm(r)$. \\

{\bf 10. Scattering phase shifts} \\
The afore mentioned geminals $R_l(r,k)$ are scattering states with an
asymptotical behavior
\begin{equation}
R_l(r,k)\to \frac{1}{kr}\sin [kr-l\frac{\pi}{2}+\eta_l(k)]
\end{equation}
defining scattering phase shifts $\eta_l(k)$ [$v_\pm(r)$ is assumed
to decay stronger than $1/r^2$]. For the case `no polarization'
the PD normalizations
\begin{equation}
\int d^3r\; \rho [1-g_p(r)]=2,\quad \int d^3r\; \rho [1-g_a(r)]=0
\end{equation}
can be equivalently reformulated as the Friedel like SR \cite{ZiePR}
\begin{equation}
\frac{2}{\pi}\sum\nolimits_L^\pm\int_0^\infty dk [-\mu'(k)]\eta_l(k)= \pm c ,
\quad \mu(k)=\frac{1}{N}\sum_\mathbf{K}n(k_1)n(k_2)
\end{equation}
with $k_{1,2}=|\frac{1}{2}\mathbf{K}\pm\mathbf{k}|$.
Generalizing this to the case of non-vanishing polarization, one should expect
e.g. for the spin-parallel PDs
\begin{equation}
\frac{2}{\pi}\sum\nolimits_L^-\int_0^\infty dk [-\mu'_{s}(k)]\eta_l(k)=
- c_{s}, \quad \mu_{s}(k)=\frac{1}{N_s}\sum_\mathbf{K}n_s(k_1)n_s(k_2)
\end{equation}
as an equivalent reformulation of the normalization (\ref{paraPD}).
$3k^2\mu_{s}(k)$ is the probability of finding two electrons with spin $s$ and
momenta $\mathbf{k}_1$ and $\mathbf{k}_2$ with the half momenta difference
$k=\frac{1}{2}|\mathbf{k}_1-\mathbf{k}_2|$ \cite{foo3}. Similar SRs should hold
for electron pairs with antiparallel spins. Also the contraction SRs (29) of
Ref. \cite{Ziepss} can be generalized to the case of non-vanishing spin
polarization. \\

{\bf Summary and outlook} \\
Based on the spin-structure of the 2-matrix [Eq.(\ref{gamma2pm})], its
contraction properties (\ref{contractionpara}) and (\ref{contractionanti})
lead to three two-variable functions
$G_{\uparrow\uparrow}(x,y)$, $G_{\downarrow\downarrow}(x,y)$, and
$G_{a}(x,y)$, which are key functions, because they contain not only the PDs
according to $g_{ss}(r)=G_{ss}(0,r)$ and $g_a(r)=G_a(0,r)$ with $g_{ss}(0)=0$,
$g_a(0)<1$ and $g_{ss}(r), g_a(r)\geq 0$, but also the 1-matrices according to
$f_s(r)=G_{ss}(r,\infty)$ with $f_s(0)=1$. These key functions $G_{ss}$ and
$G_a$ have to obey the contraction SRs (\ref{paracontraction}) and
(\ref{anticontraction2}). They can be approximately calculated (i)
perturbatively, cf. Sec. 8 and Ref. \cite{Zie5} or (ii) by the solution of an
effective 2-body equation according to the Kimball-Overhauser approach
\cite{Over, Gor3,Dav}. The Friedel like normalization and contraction SRs of
Refs. \cite{ZiePR,Ziepss} can be generalized to the case of non-vanishing
spin-polarization. The general analysis presented here has to be completed by
numerical studies. Also the question should be studied, whether conclusions
can be drawn for the afore mentioned key functions and geminals from the 
hierarchy of contracted Schr\"odinger equations \cite{Cios}, the Kummer variety 
\cite{Col1}, linear inequalities for density matrices \cite{Davi2}, the
$P/D$, $Q$, and $G$ positivity conditions \cite{Nak,Maz}. Future work should concern 
also the generalization of the fluctuation analysis of the spin-unpolarized HEG 
\cite{Zie1} to the case of non-vanishing spin-polarization. \\

{\bf Acknowledgments} \\
The authors thank P. Gori-Giorgi and J. P. Perdew for posing the question of
how to obtain the 1-matrix by contracting the 2-matrix in the case of the
spin-polarized HEG. One of the authors (P.Z.) gratefully acknowledges P. Fulde
for supporting this work. The other author (F.T.) thanks H. Eschrig for his
support of this work and the Technische Universit\"at Dresden for a 
scholarship. \\

{\bf Appendix 1: The case of `no interaction'} \\
Because of $T\sim 1/r_s^2$ and $V\sim 1/r_s$, this case is realized for
$r_s\to 0$, what effects the following simplifications, where $N_\uparrow/N=
(1+\zeta)/2$, $N_\downarrow/N=(1-\zeta)/2$ and $k_{{\rm F}\uparrow}/k_{\rm F}=
(1+\zeta)^{1/3}$, $k_{{\rm F}\downarrow}/k_{\rm F}=(1-\zeta)^{1/3}$, and
wavelengths are measured in units of $k_{\rm F}$, lengths in units of
$k_{\rm F}^{-1}$, and energies in a.u.:
\begin{equation}
n_\uparrow^{0}(k)=\theta((1+\zeta)^{1/3}-k), \quad
n_\downarrow^{0}(k)=\theta((1-\zeta)^{1/3}-k),
\label{mom0}
\end{equation}
\begin{equation}
f_\uparrow^{0}(r)=3\frac{j_1((1+\zeta)^{1/3}r)}{(1+\zeta)^{1/3}r}, \quad
f_\downarrow^{0}(r)=3\frac{j_1((1-\zeta)^{1/3}r)}{(1-\zeta)^{1/3}r},
\end{equation}
\begin{equation}
\frac{t_\uparrow^{0}}{t^{0}}=\frac{1}{2}(1+\zeta)^{2/3}, \quad
\frac{t_\downarrow^{0}}{t^{0}}=\frac{1}{2}(1-\zeta)^{2/3}, \quad
t^{0}=\frac{3}{10}\frac{1}{\alpha_0^2}, \quad \alpha_0=
\left (\frac{4}{9\pi}\right )^{1/3}r_s.
\end{equation}
There are two Fermi spheres, a large one for spin $\uparrow$ and a smaller
one for spin $\downarrow$. There are three regions of different occupancies:
$\uparrow\downarrow$ in the inner sphere, $\uparrow$ in the shell, and 0
outside the shell. Because of the idempotency $[n_s^{0}(k)]^2=n_s^{0}(k)$
and the projection $n_\uparrow^{0}(k)n_\downarrow^{0}(k)=
n_\downarrow^{0}(k)$ from Eq.(\ref{beta}) it follows
\begin{eqnarray}
\beta_{\uparrow}^{0}(r)=\frac{1+\zeta}{2}f_\uparrow^{0}(r), \quad
\beta_{\downarrow}^{0}(r)=\beta_a^{0}(r)=
\frac{1-\zeta}{2}f_\uparrow^{0}(r),
\nonumber \\
b_\uparrow^{0}=\frac{1+\zeta}{2}, \quad \quad \quad \quad
b_\downarrow^{0}=b_a^{0}=\frac{1-\zeta}{2}.
\label{beta0}
\end{eqnarray}
All the quantities and relations of Table II hold [with the replacements HF
$\to$ 0 and $f_s(r)\to f_s^{0}(r)$] also for the ideal Fermi gas including
Eq.(\ref{beta0}) and
\begin{equation}
\alpha_{3,4}^{0}(r)=\alpha^{0}(r)=\frac{1-\zeta}{2}f_\downarrow^{0}(r),
\quad \tilde \alpha^{0}(r)=0.
\end{equation}
It follows $a^{0}=(1-\zeta)/2$, what can be proved also in the following way. Starting with
the definition $a=-{\rm Tr}\gamma_a^{3,4}/N$ one has to
evaluate ${\rm Tr}\gamma_a^3$ or ${\rm Tr}\gamma_a^4$ for `no interaction',
which allows with Eq.(\ref{mom0}) to calculate the traces in the $\mathbf{k}$
space:
\begin{eqnarray}
{\rm Tr}\gamma_a^3=\sum_{\mathbf{k}_{1,2}}\langle
\hat{a}_{\downarrow\mathbf{k}_1}^{\dag}\hat{a}_{\uparrow\mathbf{k}_2}^{\dag}
\hat{a}_{\downarrow\mathbf{k}_2}^{}\hat{a}_{\uparrow\mathbf{k}_1}^{}
\rangle
=\sum_{\mathbf{k}_{1,2}}\langle
\hat{a}_{\downarrow\mathbf{k}_1}^{\dag}
\hat{a}_{\uparrow\mathbf{k}_1}^{}
\hat{a}_{\uparrow\mathbf{k}_2}^{\dag}
\hat{a}_{\downarrow\mathbf{k}_2}^{}\rangle -
\sum_{\mathbf{k}}\langle
\hat{a}_{\downarrow\mathbf{k}}^{\dag}
\hat{a}_{\uparrow\mathbf{k}}^{}
\rangle=-N_{\downarrow}.
\end{eqnarray}
Thereby $\hat{a}_{\uparrow\mathbf{k}_2}^{\dag}
\hat{a}_{\downarrow\mathbf{k}_2}^{}\rangle$ should be a state with a hole
at $\mathbf{k}_2$ in the $\downarrow$-sphere (hence $k_2<k_{F\downarrow}$)
and an additional particle at $\mathbf{k}_2$ in the $\uparrow$-sphere.
This is not possible: $\left( \hat{a}_{\uparrow\mathbf{k}}^{\dag} \right)^2=0$.
From ${\rm Tr}\gamma_a^4$ follows the same result:
\begin{eqnarray}
{\rm Tr}\gamma_a^4=\sum_{\mathbf{k}_{1,2}}\langle
\hat{a}_{\uparrow\mathbf{k}_1}^{\dag}\hat{a}_{\downarrow\mathbf{k}_2}^{\dag}
\hat{a}_{\uparrow\mathbf{k}_2}^{}\hat{a}_{\downarrow\mathbf{k}_1}^{}
\rangle
=&\sum_{\mathbf{k}_{1,2}}\langle
\hat{a}_{\uparrow\mathbf{k}_1}^{\dag}
\hat{a}_{\downarrow\mathbf{k}_1}^{}
\hat{a}_{\downarrow\mathbf{k}_2}^{\dag}
\hat{a}_{\uparrow\mathbf{k}_2}^{}\rangle -
\sum_{\mathbf{k}}\langle
\hat{a}_{\uparrow\mathbf{k}}^{\dag}
\hat{a}_{\downarrow\mathbf{k}}^{}
\rangle.
\end{eqnarray}
In the first term, with $\hat{a}_{\uparrow\mathbf{k}_2}^{}$ a hole in the
$\uparrow$-sphere can be made only in the shell $k_{F\downarrow}<k_2<
k_{F\uparrow}$ between the two Fermi surfaces.
Because $k_2$ is thus outside the $\downarrow$-sphere, the $\downarrow$ places
in the shell are empty and can be occupied:
$\hat{a}_{\downarrow\mathbf{k}_2}^{\dag}
\hat{a}_{\uparrow\mathbf{k}_2}^{}\rangle \ne 0$. With
$\mathbf{k}_1=\mathbf{k}_2$ and
$\sum_{k_{F\downarrow}<k<k_{F\uparrow}} 1=N_{\uparrow}-N_{\downarrow}$,
again $a^{0}=N_{\downarrow}/N=(1-\zeta)/2$ turns out. \\

The PDs for spin-parallel pairs are
\begin{eqnarray}
g_{\uparrow\uparrow}^{0}(r)=1-|f_\uparrow^{0}(r)|^2, \quad
\int d^3r\; \rho[1-g_{\uparrow\uparrow}^{0}(r)]=\frac{2}{1+\zeta}, \quad
g_{\uparrow\uparrow}^{0}(\infty)=1, \nonumber \\
g_{\downarrow\downarrow}^{0}(r)=1-|f_\downarrow^{0}(r)|^2, \quad
\int d^3r\; \rho[1-g_{\downarrow\downarrow}^{0}(r)]=\frac{2}{1-\zeta}, \quad
g_{\downarrow\downarrow}^{0}(\infty)=1.
\label{PDspinparallel}
\end{eqnarray}
The PDs for the spin-antiparallel pairs are
\begin{eqnarray}
g_a^{0}(r)=1, \quad &\int d^3r\; \rho[1-g_a^{0}(r)]=0,& \quad
g_a^{0}(\infty)=1, \nonumber \\
g_a'^{0}(r)=-{\rm Re}f_\uparrow^{0}(r)f_\downarrow^{0}(r), \quad
&\int d^3r\; \rho g_a'^{0}(r)=\frac{2}{1+\zeta},& \quad
g_a'^{0}(\infty)=0, \nonumber \\
g_{\pm}^{0}(r)=1\mp {\rm Re}f_\uparrow^{0}(r)f_\downarrow^{0}(r), \quad
&\int d^3r\; \rho[1-g_{\pm}^{0}(r)]=\mp\frac{2}{1+\zeta},& \quad g_{\pm}(\infty)=1.
\end{eqnarray}
Finally the spin-weighted PD is
\begin{eqnarray}
g^{0}(r)&=&\left (\frac{1+\zeta}{2}\right )^2g_{\uparrow\uparrow}^{0}+
\left (\frac{1-\zeta}{2}\right )^2g_{\downarrow\downarrow}^{0}+
\frac{1-\zeta^2}{2} \quad {\rm or} \nonumber \\
1-g^{0}(r)&=&\left (\frac{1+\zeta}{2}\right )^2|f_\uparrow^{0}(r)|^2
+\left (\frac{1-\zeta}{2}\right )^2|f_\downarrow^{0}(r)|^2
\end{eqnarray}
with $\int d^3r\; \rho[1-g^{0}(r)]=1$,
$\quad g^{0}(\infty)=1$, besides $\Delta g^{0}(r)=g_a'^{0}(r)$. \\

{\bf Appendix 2: The case of `no spin-polarization'} \\
The simplifications for `no polarization' are:
$\zeta=0$, consequently $N_\uparrow=N_\downarrow=N/2$, and also $\rho_\uparrow=
\rho_\downarrow=\rho/2$, $\gamma_\uparrow=\gamma_\downarrow=\gamma/2$,
$f_\uparrow=f_\downarrow=f$, $\gamma=\rho f$, $n_\uparrow=n_\downarrow=n$.
In this case the spin and position matrices are
\begin{eqnarray}
\delta_p  \equiv\delta_{\uparrow\uparrow}+\delta_{\downarrow\downarrow}=
\delta_{\sigma_1\sigma'_1}\delta_{\sigma_2\sigma'_2}\delta_{\sigma_1\sigma_2},
\quad\langle\uparrow\uparrow\uparrow\uparrow\rangle   =
\langle\downarrow\downarrow\downarrow\downarrow\rangle \;\; &{\rm or}&\;\;
\gamma_{\uparrow\uparrow}=\gamma_{\downarrow\downarrow}\equiv \gamma_p,
\;\; {\rm Tr}\gamma_p=\frac{N}{2}\left(\frac{N}{2}-1\right), \nonumber \\
\delta_a  \equiv\delta_a^1+\delta_a^2=
\delta_{\sigma_1\sigma'_1}\delta_{\sigma_2\sigma'_2}\delta_{\sigma_1,-\sigma_2},
\quad\langle\uparrow\downarrow\downarrow\uparrow\rangle =
\langle\downarrow\uparrow\uparrow\downarrow\rangle \;\; &{\rm or}& \;\;
\gamma_a^1=\gamma_a^2\equiv \gamma_a, \;\; {\rm Tr}\gamma_a=
\left(\frac{N}{2}\right)^2, \nonumber \\
\delta'_a  \equiv \delta_a^3+\delta_a^4=
\delta_{\sigma_1\sigma'_2}\delta_{\sigma_2\sigma'_1}\delta_{\sigma_1,-\sigma_2},
\quad \langle\downarrow\uparrow\downarrow\uparrow\rangle =
\langle\uparrow\downarrow\uparrow\downarrow\rangle \;\; &{\rm or}& \;\;
\gamma_a^3=\gamma_a^4\equiv \gamma'_a, \;\; {\rm Tr}\gamma'_a=-\frac{N}{2}.
\end{eqnarray}
An immediate consequence is ${\tilde \gamma}_a={\tilde \gamma}'_a=
{\tilde \gamma}_\pm=0$. Besides from Table I it follows
$\gamma^{\text{HF}}_a+\gamma_a^{'\text{HF}}=\gamma_p^{\text{HF}}$ and the
Feynman diagrams (42) and (44) lead correspondingly to
$\chi_a+\chi'_a=\chi_p$, because the spin-directions $\uparrow$ and $\downarrow$
are equally weighted. (Note also, that $\gamma_p$ and $\gamma_-=\gamma_a+\gamma'_a$ 
have the same symmetry property, namely changing sign under the replacements 
${\mathbf r}_1\leftrightarrow {\mathbf r}_2$ or ${\mathbf r}'_1\leftrightarrow {\mathbf r}'_2$.) 
So it is 
\begin{equation}
\gamma_a+\gamma'_a=\gamma_p \quad {\rm or} \quad \gamma'_a=\gamma_p-\gamma_a,
\quad \text{or} \quad \gamma_-=\gamma_p
\label{gammap=}
\end{equation}
(for $r_s=0$ this can be proved directly). Eq. (\ref{gammap=}) says that for $\zeta=0$ the 
2-matrix $\gamma_2$ consists of only two components,
namely $\gamma_p$ and $\gamma_a$ or $\gamma_+$ and $\gamma_-$. Indeed
the parallel/antiparallel representation (\ref{gamma2pa}) takes the
simple form
\begin{eqnarray}
\gamma_2 & = & (\delta_p+\delta'_a)\gamma_p+(\delta_a-\delta'_a)\gamma_a,
\nonumber \\
{\rm Tr}\gamma_2 & = & 2\cdot\frac{N}{2}\left (\frac{N}{2}-1\right )+
2\cdot\left ( \frac{N}{2} \right )^2= N(N-1)
\label{zeta=0}
\end{eqnarray}
and the equivalent triplet/singlet representation (\ref{gamma2pm}) simplifies as
\begin{eqnarray}
\gamma_2 &=& (\delta_p+\delta_+)\gamma_-+\delta_-\gamma_+ \nonumber \\
{\rm Tr}\gamma_2 &=& 3 \cdot\frac{N}{2}\left (\frac{N}{2}-1\right )+
1\cdot\frac{N}{2}\left (\frac{N}{2}+1\right )=N(N-1).
\label{zeta=0tripsing}
\end{eqnarray}
Its equivalence with Eq.(\ref{zeta=0}) follows from Eq.(\ref{gammap=}) and
$\gamma_a=(\gamma_++\gamma_-)/2$ and $\delta_{\pm}=(\delta_a\pm\delta_a')/2$
of Table I. The spin structure of
Eq.(\ref{zeta=0tripsing}) agrees with the result of the RPA-like treatment for
small $r_s$ \cite{Zie5}.
Note ${\rm Tr}\gamma_\pm=\frac{N}{2}(\frac{N}{2}\pm 1)$. The spin-matrices
appearing in Eqs.(\ref{zeta=0}), (\ref{zeta=0tripsing}) are
\begin{eqnarray}
\delta_p+\delta'_a=&
\delta_{\sigma_1,\sigma'_1}\delta_{\sigma_2,\sigma'_2}
\delta_{\sigma_1,\sigma_2}+
\delta_{\sigma_1,\sigma'_2}\delta_{\sigma_2,\sigma'_1}
\delta _{\sigma_1,-\sigma_2}& \to \delta_{\sigma_1,\sigma_2}, \nonumber \\
\delta_p+\delta_+=&\frac{1}{2}
(\delta_{\sigma_1,\sigma'_1}\delta_{\sigma_2,\sigma'_2}+
\delta_{\sigma_1,\sigma'_2}\delta_{\sigma_2,\sigma'_1})
&\to\frac{1}{2}(1+\delta_{\sigma_1,\sigma_2}),
\nonumber \\
\delta_-=\frac{1}{2}(\delta_a-\delta'_a)=&
\frac{1}{2}(\delta_{\sigma_1,\sigma'_1}\delta_{\sigma_2,\sigma'_2}-
\delta_{\sigma_1,\sigma'_2}\delta_{\sigma_2,\sigma'_1})
& \to \frac{1}{2}\delta_{\sigma_1,-\sigma_2}.
\end{eqnarray}
(In the last line it holds $\frac{1}{2}(\cdots)=\frac{1}{2}(\cdots)\delta_{\sigma_1,-\sigma_2}$, 
because of $1=\delta_{\sigma_1,\sigma_2}+\delta_{\sigma_1,-\sigma_2}$ and $\frac{1}{2}(\cdots)
\delta_{\sigma_1,\sigma_2}=0$.)
The arrows indicate the diagonal elements, which give the spin-dependent PD D$\gamma_2=\rho_2
=\rho^2g$ and the spin-summed PD $g(r)=\sum\limits_{\sigma_{1,2}}g(1,2)$ equivalently as
\begin{eqnarray}
g(1,2)=&\frac{1}{4}[\delta_{\sigma_1,\sigma_2}g_p(r_{12})+
\delta_{\sigma_1,-\sigma_2}g_a(r_{12})], \quad &g(r)=\frac{1}{2}[g_p(r)+g_a(r)]
\quad {\rm or} \nonumber \\
g(1,2)=&\frac{1}{4}\left [\frac{1}{2}(1+\delta_{\sigma_1,\sigma_2})g_-(r_{12})+
\frac{1}{2}\delta_{\sigma_1,-\sigma_2}g_+(r_{12})\right ], \quad
&g(r)=\frac{1}{4}[3g_-(r)+g_+(r)],
\end{eqnarray}
in agreement with Eq.(\ref{gofr}). From the above relations between the $p,a$
and the $\pm$ components of the 2-matrix follow corresponding relations for the
PD:
\begin{eqnarray}
\gamma_p=\gamma_-, \gamma=\frac{1}{2}(\gamma_++\gamma_-) \quad {\rm or} \quad 
\gamma_+=2\gamma_a-\gamma_p,\gamma_-=\gamma_p, \nonumber \\ 
g_p=g_-, g_a=\frac{1}{2}(g_++g_-) \quad {\rm or} \quad g_+=2g_a-g_p, g_-=g_p.
\end{eqnarray}
Whereas in general $\Delta g=(g_+-g_-)/2$
(= singlet-triplet PD splitting), for $\zeta=0$ it is $\Delta g=g_a-g_p$
(= antiparallel-parallel PD difference) or $g_p=g-\Delta g/2$,
$g_a=g+\Delta g/2$. So, $\Delta g$ may be addressed as magnetic PD. 
The contraction of Eq.(\ref{gammap=}) yields $\frac{\rho}{2}f(\frac{N}{2}-1)
=\frac{\rho}{2}f\frac{N}{2}-\rho \alpha$ or $\alpha(r)=f(r)/2$, thus for
$r=0$ it follows $a=1/2$. In the contraction and normalization
of the HF and cumulant parts appear $\beta_{s}=\beta_a\equiv \beta$, thus
$b_{s}=b_a\equiv b$ and $c_s=c_a\equiv c$, $b=(1-b)/2$.
The case $\zeta=0$ is summarized in Table IV.

The PDs are defined and normalized according to (with
$\mathbf{r}'_1= \mathbf{r}_1$, $\mathbf{r}'_2= \mathbf{r}_2$, $r=r_{12}$)
\begin{eqnarray}
\left (\frac{\rho}{2}\right )^2g_{\pm}(r)=
\gamma_{\pm}(\mathbf{r}_1|\mathbf{r}_1,\mathbf{r}_2|\mathbf{r}_2)&,& \;
\int d^3r\; \rho[1-g_{\pm}(r)]=\mp2,\; g_{\pm}(\infty)=1.
\end{eqnarray}
With the functions
\begin{eqnarray}
\left (\frac{\rho}{2}\right )^2G_{\pm}(r_{11'},r_{12})&=&
\int\frac{d\Omega_{12}}{4\pi}
\gamma_{\pm} (\mathbf{r}_1|\mathbf{r}'_1,\mathbf{r}_2|\mathbf{r}_2)
\end{eqnarray}
the contraction SRs are
\begin{eqnarray}
\int d^3r_{12}\; \rho[f(r_{11'})-G_{\pm}(r_{11'},r_{12})]=\mp 2f(r_{11'})&,
\quad G_{\pm}(r_{11'},\infty)=f(r_{11'}).
\label{29}
\end{eqnarray}
From the spectral resolution (\ref{specres2}) it follows
\begin{eqnarray}
G_\pm(r_{11'},r_{12})&=&\int\frac{d\Omega_{12}}{4\pi}\frac{4}{\rho^2}
\sum_{\mathbf{K},\mathbf{k}}n(|\frac{1}{2}\mathbf{K}+\mathbf{k}|)
n(|\frac{1}{2}\mathbf{K}-\mathbf{k}|)\varphi_\pm(\mathbf{r}_1,\mathbf{r}_2)
\varphi_\pm^*(\mathbf{r}'_1,\mathbf{r}_2),  \nonumber \\ 
\varphi_\pm(\mathbf{r}_1,\mathbf{r}_2)&=&\frac{1}{\sqrt\Omega}
{\rm e}^{{\rm i}\mathbf{K}\mathbf{R}}\frac{4\pi}{\sqrt\Omega}\sum_L^{\pm}
{\rm i}^lR_L(\mathbf{r},\mathbf{k}), \quad 
R_L(\mathbf{r},\mathbf{k})=R_l(r,k)Y_L(\mathbf{e}_r)Y_L^*(\mathbf{e}_k).
\end{eqnarray}
Here the functions $G_{\pm}$ contain the PDs $g_{\pm}(r)=G_{\pm}(0,r)$, and the 
1-matrix $f(r)= G_{\pm}(r,\infty)$. For `no polarization' there are the 
contraction SRs (\ref{29}) and $f(0)=1$. \\

{\bf Appendix 3. The case of `full spin polarization'} \\
Here $\zeta=1$ is considered, i.e. $N_\downarrow=0$, thus $\rho_\downarrow,
\gamma_\downarrow,f_\downarrow,n_\downarrow=0$ and $N_\uparrow=N$. The
above formulae simplify as
\begin{equation}
\gamma_1(1|1')=\delta_{\sigma_1,\sigma'_1}
\delta_{\sigma_1,\uparrow}\gamma(\mathbf{r}_1|\mathbf{r}'_1),
\gamma(\mathbf{r}_1|\mathbf{r}'_1)=\rho f(r_{11'}), f(0)=1,
\end{equation}
\begin{equation}
\gamma_2(1|1',2|2')=\delta_+(\sigma_1|\sigma'_1,\sigma_2|\sigma'_2)
\delta_{\sigma_1,\uparrow}\delta_{\sigma_2,\uparrow}
\gamma(\mathbf{r}_1|\mathbf{r}'_1,\mathbf{r}_2|\mathbf{r}'_2),
\quad{\rm Tr}\gamma_2=N(N-1),
\end{equation}
\begin{equation}
\rho^2g(r_{12})=\gamma(\mathbf{r}_1|\mathbf{r}_1,\mathbf{r}_2|\mathbf{r}_2),
\int d^3r\rho[1-g(r)]=1, \quad g(\infty)=1,
\end{equation}
\begin{equation}
\rho^2G(r_{11'},r_{12})=\int\frac{d\Omega_{12}}{4\pi}
\gamma(\mathbf{r}_1|\mathbf{r}'_1,\mathbf{r}_2|\mathbf{r}_2),
\quad G(0,r_{12})=g(r_{12}),
\end{equation}
\begin{equation}
\int d^3r_{12}\; \rho[f(r_{11'})-G(r_{11'},r_{12})]=f(r_{11'}),\quad G(r_{11'},
\infty)=f(r_{11'}).
\label{SRfull}
\end{equation}
Note $g_{\downarrow\downarrow}(r)=0$, thus $\int d^3r\; \rho
[1-g_{\downarrow\downarrow}(r)]=2/(1-\zeta)$ diverges for $\zeta\to 1$.  \\

{\bf Appendix 4: The spin structure of} $\gamma_2$\\
Eq.(\ref{gamma2pm}) agrees term by term with the Eq.(6-13) of Ref.
\cite{Davi}, where
the block-diagonality of $\gamma_2$ in the representation corresponding to the
common eigenstates of the 2-body spin-operators $\hat s_{z1}+\hat s_{z2}$ and
$(\mathbf{\hat s}_1+\mathbf{\hat s}_2)^2$:
\begin{equation}
\gamma_2=(C_{11}D_{111}C'_{11}+C_{1-1}D_{11-1}C'_{1-1}+C_{10}D_{110}C'_{10})+
C_{00}D_{000}C'_{00}+(C_{10}D_{100}C'_{00}+C_{00}D_{010}C'_{10})
\end{equation}
Here $C_{SM}$, $C'_{SM}$, $D_{SS'M}$ are abbreviations for
$C_{SM}^{\sigma_1\sigma_2}$, $C_{SM}^{\sigma'_1\sigma'_2}$,
$D_{SS'M}(\mathbf{r}_1,\mathbf{r}_2|\mathbf{r}'_1,\mathbf{r}'_2)$, respectively.
For the position matrices it is e.g. $D_{111}=\gamma_{\uparrow\uparrow}$,
$D_{11-1}=\gamma_{\downarrow\downarrow}$, $D_{110}=\gamma_{a-}$.
The triplet/singlet spin functions or (Clebsch-Gordan coefficients)
\begin{eqnarray}
C_{11}=&\delta_{\sigma_1\uparrow}\delta_{\sigma_2\uparrow}, \quad
&C_{1-1}=\delta_{\sigma_1\downarrow}\delta_{\sigma_2\downarrow}, \nonumber \\
C_{10}=&\frac{1}{\sqrt{2}}(\delta_{\sigma_1\uparrow}\delta_{\sigma_2\downarrow}+
\delta_{\sigma_2\uparrow}\delta_{\sigma_1\downarrow}), \quad
&C_{00}=\frac{1}{\sqrt{2}}(\delta_{\sigma_1\uparrow}\delta_{\sigma_2\downarrow}-
\delta_{\sigma_2\uparrow}\delta_{\sigma_1\downarrow})
\end{eqnarray}
give the corresponding spin matrices of Eqs.(\ref{spinmatrices}) and
(\ref{spinmatrices2}) according to
\begin{eqnarray}
\delta_{\uparrow\uparrow}&=C_{11}C'_{11}, \quad \delta_{\downarrow\downarrow}&=C
_{1-1}C'_{1-1},
\quad \delta_{+}=C_{10}C'_{10}, \nonumber \\
\delta_{-}&=C_{00}C'_{00}, \quad \tilde{\delta}_{+}&=C_{10}C'_{00}, \quad
\tilde{\delta}_{-}=C_{00}C'_{10}.
\end{eqnarray}

Table I. The 2-matrices of the spin-polarized HEG and their properties (cf.
Sec. 5). The
$\delta_{ss}$ and $\delta_a^1$ etc. are defined in Eq.(\ref{spinmatrices}). The
operators D, P, Q, C, and Tr are defined after Eq.(\ref{6elements}). The lines
${\uparrow\uparrow}$ and ${\downarrow\downarrow}$ are for
spin-parallel electron pairs, the following lines for spin-antiparrallel
electron pairs. The 2-matrices depend on $\mathbf{r}_1|\mathbf{r}'_1,
\mathbf{r}_2|\mathbf{r}'_2$, the PDs of column ${\rm D}\gamma$ depend on
$r_{12}=|\mathbf{r}_1-\mathbf{r}_2|$. In column ${\rm C}\gamma$ the 1-matrices
$f_\uparrow$, $f_\downarrow$, $\alpha_{3,4}$ as well as
$f= \frac{1}{2}(f_\uparrow+f_\downarrow)$,
$\tilde f=\frac{1}{2}(f_\uparrow-f_\downarrow)$,
$\alpha=\frac{1}{2}(\alpha_3+\alpha_4)$,
$\tilde \alpha =\frac{1}{2}(\alpha_3-\alpha_4)$ appear, they depend on
$r_{11'}=|\mathbf{r}_1-\mathbf{r}'_1|$. Note $f_\uparrow(0)=f_\downarrow(0)=f(0)=1$
and $\alpha_3(0)=\alpha_4(0)=\alpha(0)=a$ (which is a function
of $r_s$ and $\zeta$), hence $\tilde f(0)=0$ and $\tilde \alpha(0)=0$. \\

\begin{tabular}{|c|c||c|c|c|c|c|c|}
\hline
$\delta$ & Tr$\delta$ & $\gamma$ & D$\gamma$ & P$\gamma$ & Q$\gamma$ & C$\gamma$ & Tr $\gamma$ \\
\hline
$\delta_{\uparrow\uparrow}$ & 1 & $\gamma_{\uparrow\uparrow}=\langle\uparrow\uparrow\uparrow\uparrow\rangle$ &
$\rho_\uparrow^2g_{\uparrow\uparrow}$ & $-\gamma_{\uparrow\uparrow}$ &
$\gamma_{\uparrow\uparrow}$ & $\rho_\uparrow f_\uparrow(N_\uparrow-1)$ &
$N_\uparrow(N_\uparrow-1)$ \\
$\delta_{\downarrow\downarrow}$ & 1 & $\gamma_{\downarrow\downarrow}=
\langle\downarrow\downarrow\downarrow\downarrow\rangle$ &
$\rho_\downarrow^2g_{\downarrow\downarrow}$ & $-\gamma_{\downarrow\downarrow}$ &
$\gamma_{\downarrow\downarrow}$ & $\rho_\downarrow f_\downarrow(N_\downarrow-1)$ &
$N_\downarrow(N_\downarrow-1)$ \\
\hline
$\delta_a^1$ & 1 & $\gamma_a^1=\langle\uparrow\downarrow\downarrow\uparrow\rangle$ &
$\rho_\uparrow\rho_\downarrow g_a$ & $-\gamma_a^4$ & $\gamma_a^1$ &
$\rho_\uparrow f_\uparrow N_\downarrow$ & $N_\uparrow N_\downarrow$ \\
$\delta_a^2$ & 1 & $\gamma_a^2=\langle\downarrow\uparrow\uparrow\downarrow\rangle$ &
$\rho_\downarrow\rho_\uparrow g_a$ & $-\gamma_a^3$ & $\gamma_a^2$ &
$\rho_\downarrow f_\downarrow N_\uparrow$ & $N_\downarrow N_\uparrow $ \\
$\delta_a^3$ & 0 & $\gamma_a^3=\langle\downarrow\uparrow\downarrow\uparrow\rangle$ &
$-\rho_\uparrow\rho_\downarrow \Delta g$ & $-\gamma_a^2$ & $\gamma_a^4$ &
$-\rho\alpha_3 $ & $-Na $ \\
$\delta_a^4$ & 0 & $\gamma_a^4=\langle\uparrow\downarrow\uparrow\downarrow\rangle$ &
$-\rho_\downarrow\rho_\uparrow \Delta g$ & $- \gamma_a^1$ & $\gamma_a^3$ &
$-\rho\alpha_4 $ & $-Na $ \\
\hline
$\delta_a=\delta_a^1+\delta_a^2$& 2 & $\gamma_a=\frac{1}{2}(\gamma_a^1+\gamma_a^2)$ &
$\rho_\uparrow \rho_\downarrow g_a$ & $- \gamma'_a$ & $\gamma_a$ &
$\rho_\uparrow f N_\downarrow $ & $N_\uparrow N_\downarrow $ \\
$\delta'_a=\delta_a^3+\delta_a^4$ & 0 & $\gamma'_a=\frac{1}{2}(\gamma_a^3+\gamma_a^4)$ &
$-\rho_\downarrow\rho_\uparrow \Delta g$ & $- \gamma_a$ & $\gamma'_a$ &
$-\rho\alpha $ & $-Na $ \\
${\tilde \delta}_a=\delta_a^1-\delta_a^2$ & 0 & $\tilde \gamma_a=\frac{1}{2}(\gamma_a^1-\gamma_a^2)$ &
$0$ & $+ \tilde \gamma'_a$ & $\tilde \gamma_a$ &
$\rho_\downarrow \tilde f N_\uparrow $ & $0$ \\
${\tilde \delta}'_a=\delta_a^3-\delta_a^4$ & 0 & $\tilde \gamma'_a=\frac{1}{2}(\gamma_a^3-\gamma_a^4)$ &
$0$ & $+\tilde \gamma_a$ & $-\tilde \gamma'_a$ &
$-\rho\tilde \alpha $ & $0$ \\
\hline
$\delta_-=\frac{1}{2}(\delta_a-\delta'_a)$ & 1 & $\gamma_{+}=\gamma_a-\gamma'_a$ &
$\rho_\uparrow\rho_\downarrow(g_a+\Delta g)$ & $+\gamma_{+}$ &
$\gamma_{+}$ & $\rho_\uparrow f N_\downarrow+\rho\alpha$ &
$N_\uparrow N_\downarrow+Na$ \\
$\delta_+=\frac{1}{2}(\delta_a+\delta'_a)$ & 1 & $\gamma_{-}=\gamma_a+\gamma'_a$ &
$\rho_\downarrow\rho_\uparrow(g_a-\Delta g)$ & $-\gamma_{-}$ &
$\gamma_{-}$ & $\rho_\downarrow f N_\uparrow-\rho\alpha$ &
$N_\downarrow N_\uparrow-Na$ \\
${\tilde \delta}_-=\frac{1}{2}({\tilde \delta}_a+{\tilde \delta}'_a)$ & 0 & $\tilde \gamma_{+}=\tilde \gamma_a+\tilde \gamma'_a$ &
$0$ & $+\tilde \gamma_{+}$ &
$\tilde \gamma_{-}$ & $\rho_\uparrow \tilde f N_\downarrow-\rho\tilde \alpha$ &
$0$ \\
${\tilde \delta}_+=\frac{1}{2}({\tilde \delta}_a-{\tilde \delta}'_a)$& 0 &
$\tilde \gamma_{-}=\tilde \gamma_a-\tilde \gamma'_a$ &
$0$ & $-\tilde \gamma_{-}$ &
$\tilde \gamma_{+}$ & $\rho_\downarrow \tilde f N_\uparrow+\rho\tilde \alpha$ &
$0$ \\
\hline
\end{tabular}
\\

\newpage
Table II. The HF parts of the cumulant expansion (cf. Sec. 8). Note
$\alpha_{3,4}^{\rm HF}(r)=\alpha^{\rm HF}(r)=\beta_a^{\rm HF}$(r), $a^{\rm HF}=
b_a^{\rm HF}$, $\tilde \alpha^{\rm HF}=0$. For `no interaction' the cumulant
parts vanish, only the HF parts remain, which simplify furthermore as
$\beta_{s}^{0}(r)=\frac{N_s}{N}f_s^{0}$(r), $\beta_a^{0}(r)=
\frac{N_\downarrow}{N}f_\downarrow^{0}(r)$, thus $b_{s}^{0}=
\frac{N_s}{N}$, $b_a^{0}=\frac{N_\downarrow}{N}$.  \\

\begin{tabular}{|c|c|c|}
\hline
$\gamma^{\rm HF}$ & C$\gamma^{\rm HF}$ & Tr$\gamma^{\rm HF}$ \\
\hline
$\gamma_{ss}^{\rm HF}=\rho_s^2[f_s(r_{11'})f_s(r_{22'})
-f_s(r_{12'})f_s(r_{21'})]$ &
$\rho_sf_sN_s-\rho \beta_{s}$ & $N_s^2-Nb_{s}$ \\
\hline
$\gamma_{a}^{{1\rm HF}} =
\rho_\uparrow\rho_\downarrow f_\uparrow(r_{11'})f_\downarrow(r_{22'})$ &
$\rho_\uparrow f_\uparrow N_\downarrow$ & $N_\uparrow N_\downarrow$ \\
$\gamma_{a}^{{2\rm HF}}=
\rho_\downarrow\rho_\uparrow f_\downarrow(r_{11'})f_\uparrow(r_{22'})$ &
$\rho_\downarrow f_\downarrow N_\uparrow$ & $N_\downarrow N_\uparrow$ \\
$\gamma_{a}^{{3\rm HF}} =
-\rho_\uparrow\rho_\downarrow f_\uparrow(r_{12'})f_\downarrow(r_{21'})$ &
$-\rho \beta_a$ &  $-Nb_a$ \\
$\gamma_{a}^{{4\rm HF}}=
-\rho_\downarrow\rho_\uparrow f_\downarrow(r_{12'})f_\uparrow(r_{21'})$ &
$-\rho\beta_a$ & $-Nb_a$ \\
\hline
$\gamma_a^{\rm HF}=\frac{1}{2}\rho_\uparrow\rho_\downarrow
[f_\uparrow(r_{11'})f_\downarrow(r_{22'})+
f_\downarrow(r_{11'})f_\uparrow(r_{22'})]$ & $\rho_\uparrow f N_\downarrow$ &
$N_\uparrow N_\downarrow$ \\
${\gamma'_a}^{\rm HF}=-\frac{1}{2}\rho_\uparrow\rho_\downarrow
[f_\uparrow(r_{12'})f_\downarrow(r_{21'})+
f_\downarrow(r_{12'})f_\uparrow(r_{21'})]$ & $-\rho \beta_a$ & $-N b_a$ \\
$\tilde {\gamma_a}^{\rm HF}=\frac{1}{2}\rho_\downarrow\rho_\uparrow
[f_\uparrow(r_{11'})f_\downarrow(r_{22'})-
f_\downarrow(r_{11'})f_\uparrow(r_{22'})]$ &
$\rho_\downarrow \tilde f N_\uparrow$ & $0$ \\
$\tilde {\gamma'_a}^{\rm HF}=-\frac{1}{2}\rho_\uparrow\rho_\downarrow
[f_\uparrow(r_{12'})f_\downarrow(r_{21'})-
f_\downarrow(r_{12'})f_\uparrow(r_{21'})]$ & $0$ & $0$ \\
\hline
$\gamma_{+}^{\rm HF}=\gamma_a^{\rm HF}-{\gamma'_a}^{\rm HF}$ &
$\rho_\uparrow f N_\downarrow+\rho \beta_a$ & $N_\uparrow N_\downarrow+Nb_a$ \\
$\gamma_{-}^{\rm HF}=\gamma_a^{\rm HF}+{\gamma'_a}^{\rm HF}$ &
$\rho_\downarrow f N_\uparrow-\rho\beta_a$ & $N_\downarrow N_\uparrow-Nb_a$ \\
$\tilde \gamma_{+}^{\rm HF}=
\tilde {\gamma_a}^{\rm HF}+\tilde {\gamma'_a}^{\rm HF}$ &
$\rho_\uparrow\tilde f N_\downarrow$ & $0$ \\
$\tilde \gamma_{-}^{\rm HF}=
{\tilde \gamma_a}^{\rm HF}-\tilde {\gamma'_a}^{\rm HF}$ &
$\rho_\downarrow \tilde f N_\uparrow$ & $0$ \\
\hline
\end{tabular}
\\

\newpage
Table III. The cumulant 2-matrices, being size-extensively normalized (cf.
Sec. 8). The index `c' means `connected' or `cumulant'. The cumulant
normalization ${\rm Tr}\chi_{ss}/N_s=c_{s}$ measures the correlation strength,
cf. Eq.(\ref{Low}). \\

\begin{tabular}{|c|c|c|}
\hline
$\chi=\gamma^{\rm HF}-\gamma$ & C$\chi$ & Tr$\chi$ \\
\hline
$\chi_{ss}=-\langle ssss\rangle_{\rm c}$ &
$\rho_s f_s-\rho \beta_{s}$ & $N_s-Nb_{s}$ \\
\hline
$\chi_a^1=-\langle \uparrow\downarrow\downarrow\uparrow\rangle_{\rm c}$ &
$0$ & $0$ \\
$\chi_a^2=-\langle \downarrow\uparrow\uparrow\downarrow\rangle_{\rm c}$ &
$0$ & $0$ \\
$\chi_a^3=-\langle \downarrow\uparrow\downarrow\uparrow\rangle_{\rm c}$ &
$\rho(\alpha_3-\beta_a)$ & $N(a -b_a)$ \\
$\chi_a^4=-\langle \uparrow\downarrow\uparrow\downarrow\rangle_{\rm c}$ &
$\rho(\alpha_4-\beta_a)$ & $N(a-b_a)$ \\
\hline
$\chi_a=\frac{1}{2}(\chi_a^1+\chi_a^2)$ & $0$ & $0$ \\
$\chi'_a=\frac{1}{2}(\chi_a^3+\chi_a^4)$ & $\rho(\alpha-\beta_a)$ & $N(a-b_a)$
\\
$\tilde\chi_a=\frac{1}{2}(\chi_a^1-\chi_a^2)$ & $0$ & $0$ \\
$\tilde \chi'_a=\frac{1}{2}(\chi_a^3-\chi_a^4)$ & $\rho\tilde \alpha$ & $0$ \\
\hline
$\chi_{+}=\chi_a-\chi'_a$ & $-\rho(\alpha-\beta_a)$ & $-N(a-b_a)$ \\
$\chi_{-}=\chi_a+\chi'_a$ & $+\rho(\alpha-\beta_a)$ & $+N(a-b_a)$ \\
$\tilde \chi_{+}=\tilde \chi_{a}+\tilde \chi'_{a}$ & $+\rho\tilde \alpha$ & $0$ \\
$\tilde \chi_{-}=\tilde \chi_{a}-\tilde \chi'_{a}$ & $-\rho\tilde \alpha$ & $0$ \\
\hline
\end{tabular}
\\

Table IV. This is Table I, simplified for the case of the spin-unpolarized HEG.
In App. 2 it is shown : $\gamma_p=\gamma_{-}=\gamma_a-\gamma'_a$, thus
also $\Delta g=g_a-g_p$, and $\alpha=f/2$, $a=1/2$.
Besides, the HF parts contain $\beta_{s}=\beta_a\equiv \beta$, thus
$b_{s}=b_a\equiv b$, cf. Eqs.(\ref{beta}), (\ref{Low}). \\

\begin{tabular}{|c|c|c|c|c|c|}
\hline
$\gamma$ & D$\gamma$ & P$\gamma$ & Q$\gamma$ & C$\gamma$ & Tr$\gamma$\\
\hline
$\gamma_p=\langle\uparrow\uparrow\uparrow\uparrow\rangle$ &
$(\frac{\rho}{2})^2g_p$ & $-\gamma_p$ & $\gamma_p$ &
$\frac{\rho}{2}f (\frac{N}{2}-1)$ & $\frac{N}{2}(\frac{N}{2}-1)$ \\
\hline
$\gamma_a=\langle\uparrow\downarrow\downarrow\uparrow\rangle$ &
$(\frac{\rho}{2})^2g_a$ & $-\gamma'_a$ & $\gamma_a$ &
$\frac{\rho}{2}f \frac{N}{2}$ & $(\frac{N}{2})^2$ \\
$\gamma'_a=\langle\downarrow\uparrow\downarrow\uparrow\rangle$ &
$-(\frac{\rho}{2})^2\Delta g$ & $-\gamma_a$ & $\gamma'_a$ &
$-\frac{\rho}{2}f $ & $-\frac{N}{2}$ \\
\hline
$\gamma_{+}=\gamma_a-\gamma'_a$ & $(\frac{\rho}{2})^2(g_a+\Delta g)$ &
$+\gamma_{+}$ & $\gamma_{+}$ & $\frac{\rho}{2}f(\frac{N}{2}+1)$ &
$\frac{N}{2}(\frac{N}{2}+1)$ \\
$\gamma_{-}=\gamma_a+\gamma'_a$ & $(\frac{\rho}{2})^2(g_a-\Delta g)$ &
$-\gamma_{-}$ & $\gamma_{-}$ & $\frac{\rho}{2}f(\frac{N}{2}-1)$ &
$\frac{N}{2}(\frac{N}{2}-1)$ \\
\hline
\end{tabular}

\newpage
{\bf Captions of the Figures} \\

Fig.1. Dimensionless 1-matrix $f_s(r)$ of the spin-polarized HEG for different
values of the spin-polarization $\zeta$ and the interaction strength $r_s$.
$r$ in units of $k_{\rm F}^{-1}$. Upper panel: ideal Fermi gas, $s=\uparrow$,
and increasing $\zeta$. Middle panel: ideal Fermi gas, $\zeta=1/3$, and
$s=\uparrow,\downarrow$. Lower panel: $\zeta=1/3$, $s=\uparrow$, and
increasing $r_s$. \\

Fig.2. Momentum distributions $n_s(k,r_s,\zeta)$ of the spin-polarized HEG
according to Eq.(\ref{approx}) for the spin-polarization $\zeta=1/3$ and the
interaction-strength values $r_s=1, 5, 10$. $k$ in units of $k_{\rm F}$. \\

Fig.3. Kinetic correlation energy $t_{\rm corr}(r_s)$ in a.u. according to
Perdew/Wang \cite{Per} (full line) and from $n_s(k,r_s,\zeta)$ of
Eq.(\ref{approx}) via $t=t_\uparrow+t_\downarrow$ (the dots). \\

Fig.4. Non-idempotency matrices $\beta_s(r)$ and $\beta_a(r)$ of the
spin-polarized HEG according to Eq.(\ref{beta}). $r$ in units of
$k_{\rm F}^{-1}$. 1st and 2nd panel: ideal Fermi gas, 3rd and 4th panel:
increasing interaction strength $r_s$. \\
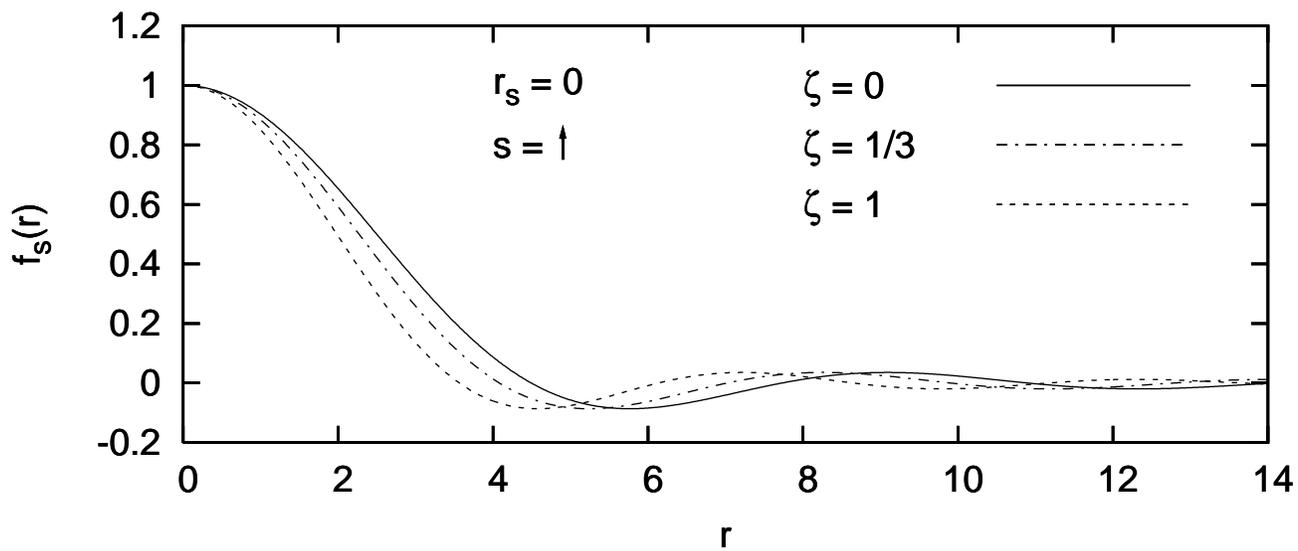
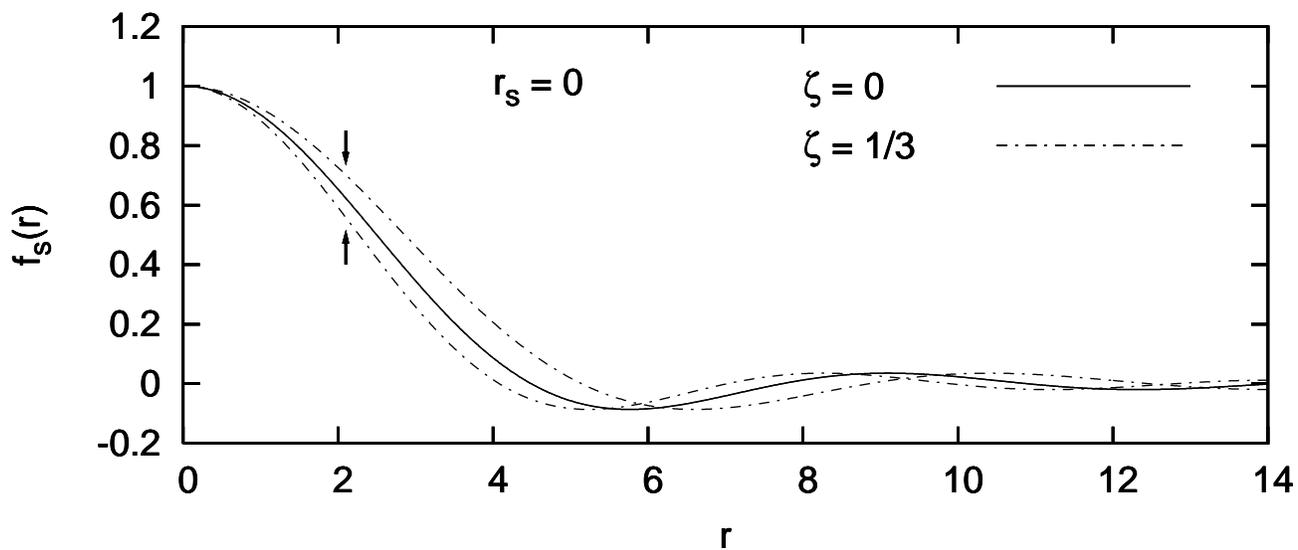
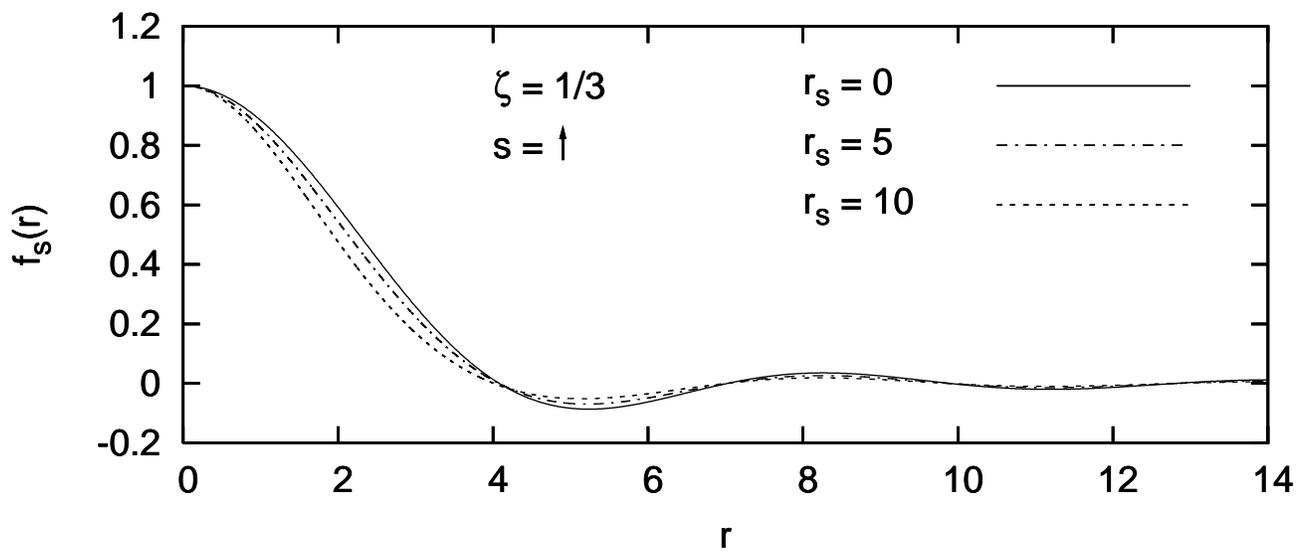
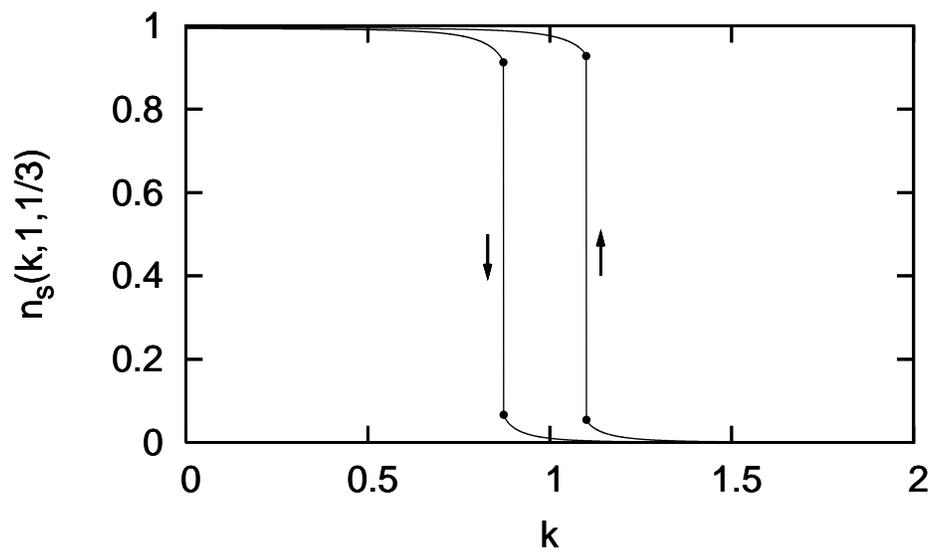
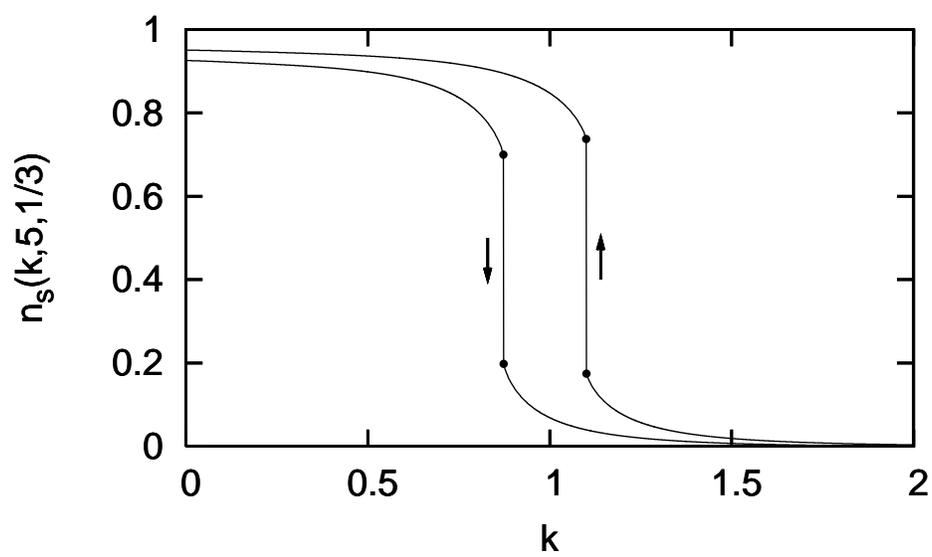
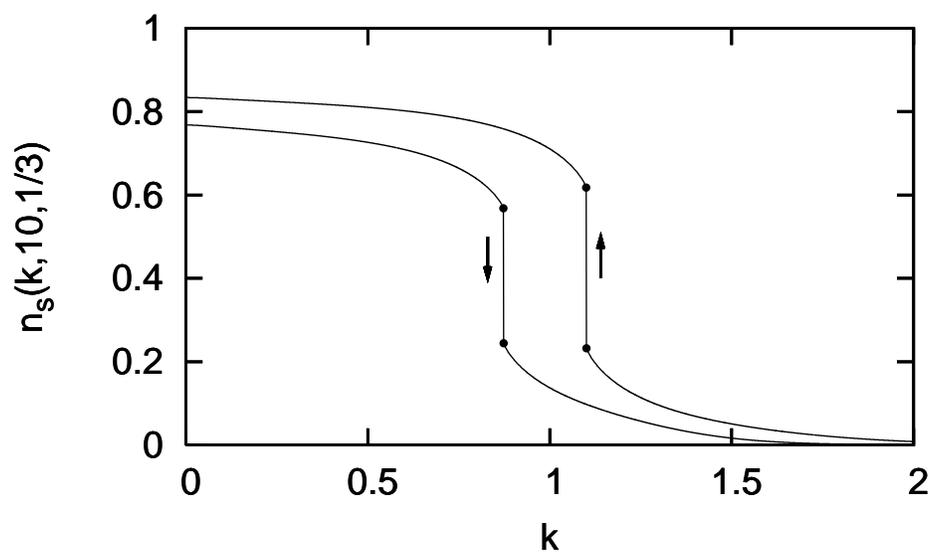
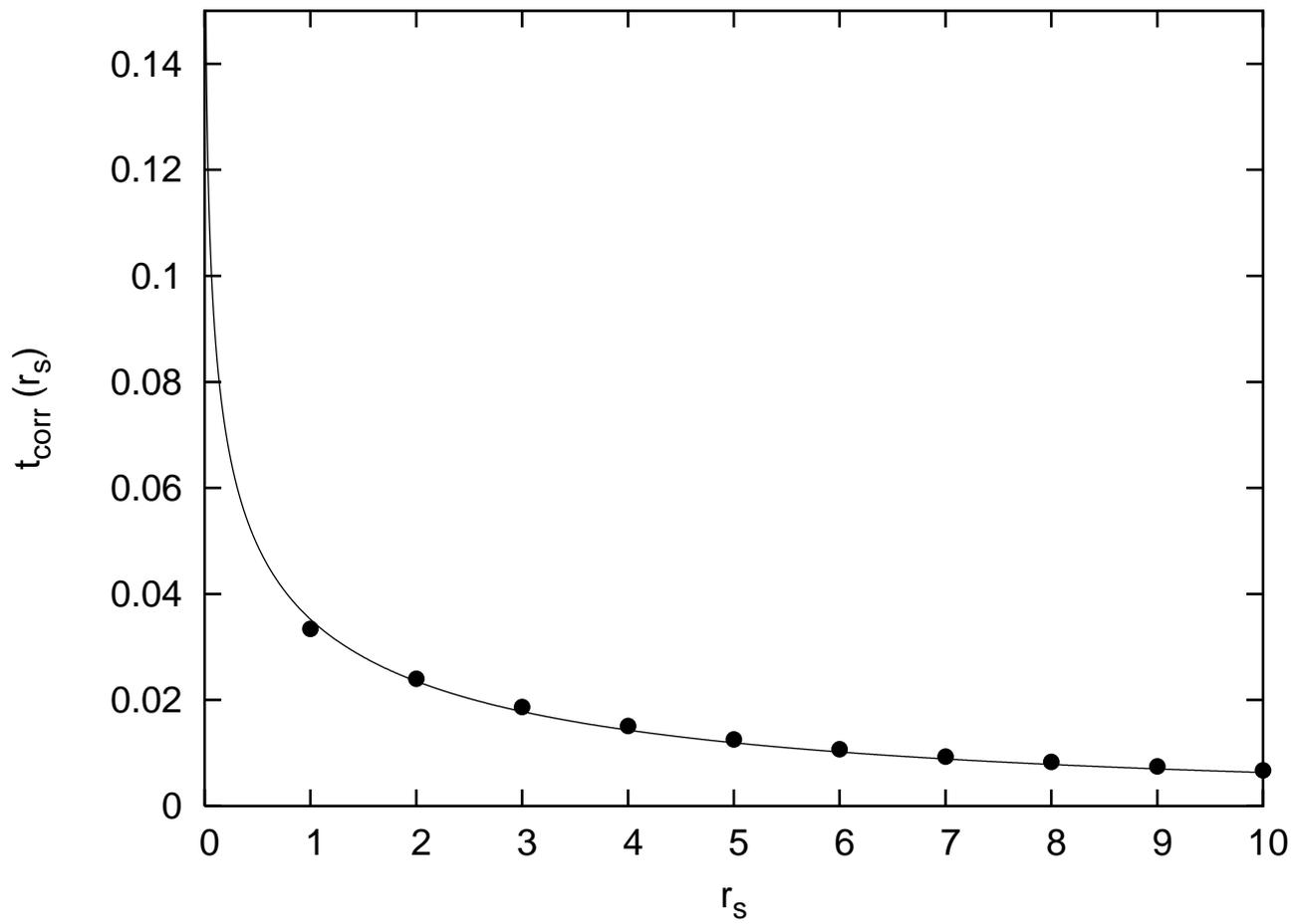
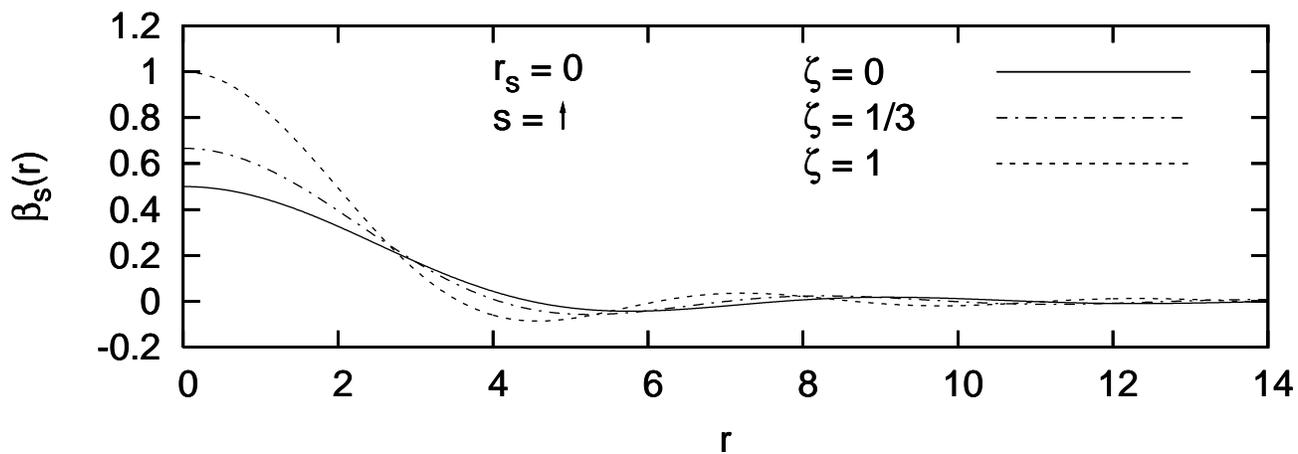
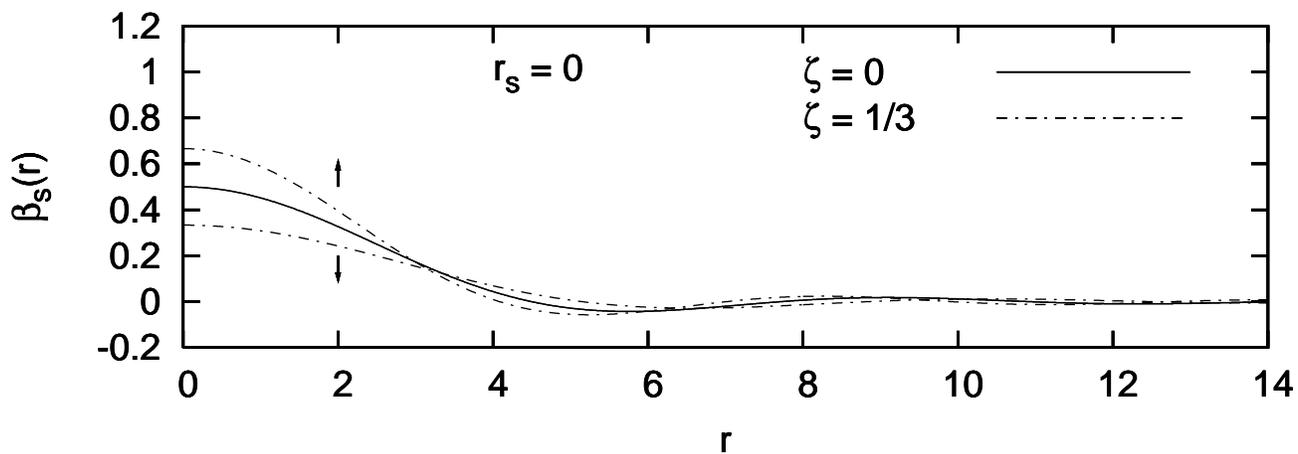
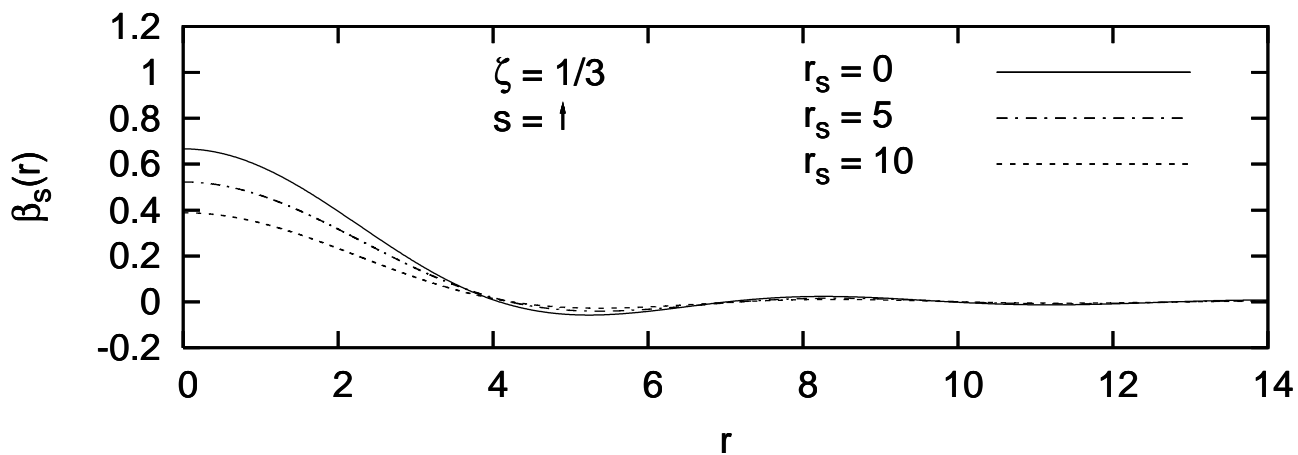
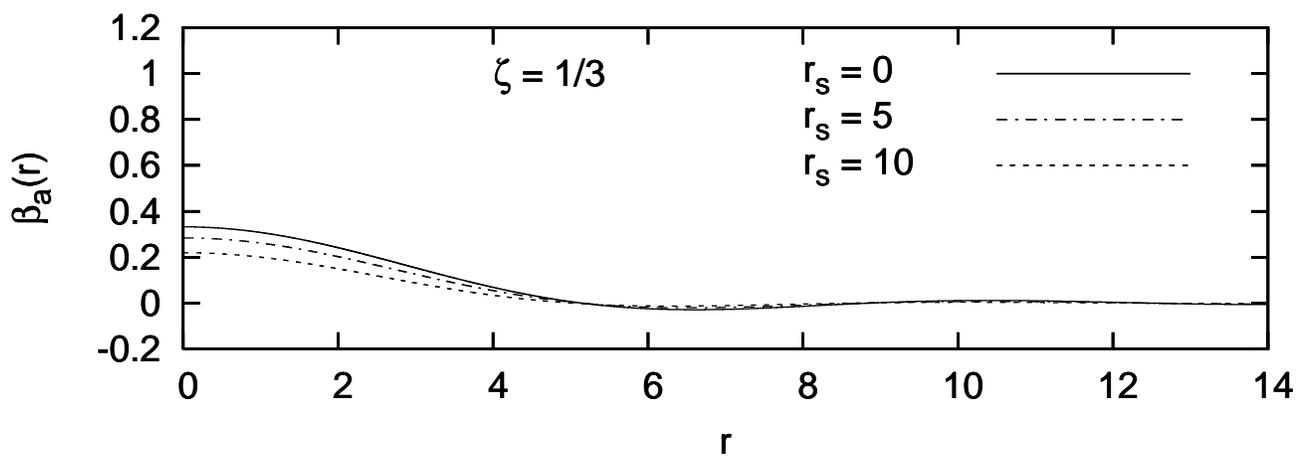
\end{document}